\documentclass[11pt]{article}
\usepackage{jheppubmod}
\usepackage{pstool}
\pdfoutput=1
\usepackage{color}
\usepackage{float}
\usepackage{array}
\usepackage{amssymb}
\usepackage{amsmath}
\usepackage{amsthm}
\usepackage{graphicx}
\usepackage{caption}
\usepackage[labelsep=quad]{subcaption}
\usepackage{epstopdf}
	
\usepackage{epsfig} 

\def\pb[#1,#2]{\{#1, #2\}}
\def\deb[#1,#2]{[#1,#2]_{\text{D.B.}}}

\def\half{{1 \over 2}}

\def\Or[#1]{{\text{O}}\left({#1}\right)}
\def\dotl[#1,#2]{\left\langle #1,\, #2 \right\rangle}
\def\dotlb[#1,#2]{\left\langle #1,\, #2 \right\rangle}
\def\dotlm[#1,#2]{\left[ #1,\, #2 \right]}
\def\dotp[#1,#2]{(\vect{#1} \cdot\vect{#2})}
\def\aff[#1,#2]{\hat{#1}(#2)}

\def\n4sym{{\cal N}=4 SYM}
\def\>{\rangle}
\def\<{\langle}
\def\weight[#1,#2,#3]{\{(#1),#2,#3\}}
\def\ads[#1]{$\text{AdS}_{#1}$}

\newcommand{\be}{\begin{equation}}
\newcommand{\ee}{\end{equation}}
\newcommand{\ba}{\begin{align}}
\newcommand{\ea}{\end{align}}
\newcommand{\bs}{\begin{split}}
	\def\sess\end{split}
\newcommand{\vect}[1]{{\boldsymbol{#1}}}

\def\g{\gamma}

\def \bea {\begin{eqnarray}}
\def \eea {\end{eqnarray}}
\def \bea* {\begin{eqnarray*}}
	\def \eea* {\end{eqnarray*}}

\def \be {\begin{equation}}
\def \ee {\end{equation}}
\def \bes {\begin{equation*}}
\def \ees {\end{equation*}}

\title{On the universality of late-time correlators in semi-classical 2d CFTs}
\author[a]{Souvik Banerjee,}
\emailAdd{souvik.banerjee@physics.uu.se}
\affiliation[a]{Department of Physics and Astronomy, Uppsala University, \\
 SE-751 08 Uppsala, Sweden\\} 
\author[b]{Jan-Willem Bryan,}
\affiliation[b]{Van Swinderen Institute for Particle Physics and Gravity, \\ University of Groningen, Nijenborgh 4, 9747 AG, The Netherlands \\}
\emailAdd{j.w.a.brijan@rug.nl}
\author[b]{and Gideon Vos}
\emailAdd{g.vos@rug.nl}

\date{}

\abstract{In the framework of AdS$_3$/ CFT$_2$ correspondence, we present a systematic analysis of the late time thermalization of a two dimensional CFT state created by insertion of small number of heavy operators on the vacuum. We show that at late Lorentzian time, the universal features of this thermalization are solely captured by the eigenvalues of the monodromy matrix corresponding to the solutions of the uniformization equation. We discuss two different ways to extract the monodromy eigenvalues while bypassing the need for finding explicitly the full monodromy matrix  - first, using a monodromy preserving diffeomorphism and second using Chen-Simons formulation of gravity in AdS$_3$. Both of the methods yield the same precise relation between the eigenvalues and the final black hole temperature at late Lorentzian time.
}
\begin{document}
\preprint{UUITP-22/18}
\maketitle
	
\section{Introduction}	
Over the last few decades black holes have been the primary window into the manifestation of quantum effects in gravity. In particular a huge amount of research in the last few years has been devoted to the understanding of the conflict between black hole physics and quantum mechanical expectation in the light of AdS/CFT correspondence \cite{Maldacena:1997re}. 

An ideal set up to address this problem is the quantum gravity theory in three-dimensional Anti-de Sitter spacetime. The dual field theory being a two dimensional CFT, it becomes easier, technically, to do explicit field theory computations exploiting the Virasoro algebra.  Furthermore, qualitatively the static and stationary black hole solution in AdS$_3$, namely the BTZ black hole \cite{Banados:1992wn}, shares most of its qualitative features with higher dimensional AdS black holes and hence one expects to get a lot of insights on the black hole information problem solely from the study of black hole physics in AdS$_3$\footnote{One might wonder that the black holes in AdS$_2$ might be even easier technically to study. However, AdS$_2$ spacetime has its own pathologies \cite{Maldacena:1998uz}. It does not support any finite energy excitations keeping the asymptotic properties fixed. Nonetheless, there has been a surge of work in understanding information loss paradox with simple one dimensional quantum mechanical toy models with nearly AdS$_2$ duals \cite{Maldacena:2016hyu, Maldacena:2016upp}. We hope to have a more complete understanding of the AdS$_2$ story in this context in near future.} 

Black holes are identified as localized high energy states in a quantum gravity theory. Therefore in the framework of AdS$_3$/ CFT$_2$ correspondence, it would be tempting to create a heavy state in CFT and understand under which conditions the late time behavior of the correlators evaluated on the state are indistinguishable to the expected thermal correlators in the black hole spacetime. In principle, from the eigenstate thermalization hypothesis \cite{srednicki1999approach} this late time thermal behavior of high energy states is expected for any dynamical complex system. However, one important feature of black hole thermalization is that it is characterized by a particular temperature, namely the Hawking temperature which is a purely geometric quantity. It would therefore be challenging to identify the quantity in the dual CFT construction which captures this universality. This is one of the main goals of this work.

\vskip10pt

\noindent Our starting point will be to construct a heavy state dual to a collapsing black hole in AdS$_3$. Our initial state would correspond to arbitrary mass distributions in the bulk in the form of some random distribution of conical defects or even colliding black holes. This CFT construction has its application beyond the understanding of the thermalization properties of black holes. This is in fact the set up where addressing the information loss paradox is most challenging. We will however take the first small step towards this in this work where we would contemplate on the collapse of this arbitrarily heavy state into a thermal background and measures to identify the collapse from a purely CFT perspective.  

The first study in this line was done in \cite{Anous:2016kss}  where the authors constructed a collapse state in CFT$_2$ by insertion of a large $n$ number of local primary operators corresponding to a large number of dust particles constituting a massive null shell in AdS$_3$. Other work in this direction include \cite{Chen:2016dfb, Chen:2016kyz}. In the limit $n \rightarrow \infty$ this yields a state with a well-defined uniform stress-energy tensor. The simple form of the ``averaged'' stress-energy tensor simplifies the computations of correlation functions drastically as demonstrated in their work. They specifically considered CFTs with large central charge, $c$ and having a sparse spectrum at low energy, the gravity dual of which corresponds to having an AdS$_3$ spacetime with a very large radius compared to the Planck length $c \sim \frac{{\cal R}_{\text AdS}}{L_{\text P}}$. For such CFTs the collapse state was constructed and the correlation functions were studied as a perturbative expansion in $\frac{1}{c}$. 

We however adopt an alternative and a more generic scenario where a black hole in AdS$_3$ is created through collision of a finite number of heavy particles. In the perspective of the dual CFT$_2$ the possible dual set up is the creation of a high energy state through insertion of primary operators with high scaling dimension. For this dual picture to make sense the scaling dimension of the inserted operators need to scale with $c$ in the limit $c \rightarrow \infty$. Furthermore, in view of having a static black hole as the final state of the time-evolution, every pair of operators need to be inserted at antipodal points with respect to one another. In our work, we however, consider more general black hole states starting with arbitrary initial energy configurations. While studying the problem numerically we will consider two separate cases, namely operator insertions resulting in a static black hole or an oscillating final black hole state. 

Once the background is created, the immediate interest will be to understand the measures of the black hole collapse at late Lorentzian time. It turns out that the most natural and simplest detector of the heavy state collapsing into a thermal black hole state is the correlation functions of the form
\begin{equation}
\mathcal{A}_{N_Q}=\langle V|Q(z_1)Q(z_2) \cdots Q(z_{N_Q})|V\rangle,
\label{correlatorN}
\end{equation}
where $|V\rangle$ is the heavy state created by insertion of heavy primaries as discussed above. The $Q(z_i)$'s are probe operators which possess the property that the scaling dimensions of these operators do not scale with $c$ in the  in the limit $c \rightarrow \infty$. These probe operators, typically termed as ``light operators'' therefore do not back-react on the heavy state. A heavy state created through insertion of $N_H$ number of heavy primaries, \eqref{correlatorN} amounts to evaluating a $\left(2 N_H + N_Q\right)$ point correlation function in the vacuum state. This correlator can be evaluated using conformal block techniques and as we will argue, the dominant contribution to the correlator comes from the vacuum Virasoro block which effectively captures the pure gravitational interactions in the bulk. This fact greatly simplifies the computation of the correlation function, \eqref{correlatorN} . However, if $N_H$ is a finite number of heavy operators naively it looks like the perturbative expansion in $\frac{1}{c}$  breaks down and it becomes  hard to extract even the leading order semi-classical behavior of the correlators, \eqref{correlatorN}. 

In a series of work initiated with \cite{Fitzpatrick:2015zha} and later developed further in \cite{Fitzpatrick:2015dlt}, the authors proposed a trick to save the perturbative expansion in $\frac{1}{c}$. Their prescription was to put the CFT$_2$ on a non-trivial background geometry. However, in $d=2$ any geometry is related to the flat space geometry by a Weyl rescaling. The proposal was to choose an appropriate Weyl scaling or equivalently, a conformal transformation, $z \rightarrow w\left(z\right)$ such that in the new coordinates, $w$, it becomes possible to write down the correlator as a power series expansion in $\frac{1}{c}$ again.

This procedure of resumming the divergences in the perturbation series at large $c$, known as the ``uniformization problem'' in the literature, is mathematically equivalent of sewing local coordinate patches around the punctures of a n-punctured Riemann sphere to construct a smooth covering manifold. However, the actual implementation of the mechanism increases in difficulty as the number of punctures increases. In the language of the correlation functions, this number denotes the number of heavy operator insertions required to create the heavy background state. 

For our case we need at least four heavy operators (two heavy operators and their adjoints) to create a static black hole configuration at late time which amounts to solving the ``uniformization problem" for a four-punctured Riemann sphere. Unfortunately, this problem does not have an analytic solution and therefore it is not possible to write down the conformal transformation $z \rightarrow w\left(z\right)$ for this case \footnote{We have found some literature like \cite{Hadasz:2006rb} where the authors made an attempt to glue different coordinate patches defined locally, in the vicinity of the punctures of a four-punctured Riemann sphere.  They used local series solutions of the uniformization equation near each puncture and finally matched the series solution at intermediate points. However, the variables they used had limited range of validity and the local series expansions, limited radius of convergences, both the facts not very useful for our purpose.}. 

\vskip10pt

\noindent 
The main claim of our work is that as long as one is concerned with the late time behavior of the Lorentzian correlator \eqref{correlatorN}, one can actually get around this problem of finding explicit solutions to the uniformization problem which in our case amounts to solving a second order differential equation of Fuchsian class. We show that most of the relevant information about the late time thermalization is already encoded in the monodromy matrix of the solutions along a curve on the unit circle. In particular the temperature of the final state is related to the eigenvalue of the monodromy matrix. However, finding the monodromy matrix explicitly for a general Fuchsian equation is technically as daunting as solving the original uniformization problem. But in this work we introduce two completely different methods to establish the precise connection between the eigenvalues of the monodromy matrix at late Lorentzian time and the final black hole temperature even without knowing the explicit matrix.

The first approach relies on finding a conformal transformation that preserves the structure of the monodromy matrix while leading to a new Fuchs equation which is much easier to solve. In fact this transformation leads to a stress-energy wave function similar to one obtained in the continuum limit of \cite{Anous:2016kss} discussed earlier.  For the hyperbolic class of monodromy matrices, this yields a precise relation between the eigenvalue of the monodromy matrix and the Hawking temperature of the black hole in the late Lorentzian time limit. 

The second approach exploits the the representation of the monodromy matrix in terms of a path ordered integral over a flat connection \cite{Castro:2013lba}. We then use the Chern-Simons formulation of pure gravity to relate this path ordered integral to the area of horizon of the black hole. Upon using the Bekenstein-Hawking formula this yields the same relation between the eigenvalue of the monodromy matrix and the Hawking temperature of the black hole after collapse. 

The first of our proofs tells us something more about the final state of the Lorentzian time evolution. In fact, at late Lorentzian time, it is also possible to obtain a black hole dressed with soft  gravitational hair. This soft hair corresponds to ``boundary gravitons'', namely  some non-propagating graviton degrees of freedom localized near the asymptotic boundary of the AdS$_3$ spacetime. From the perspective of dual CFT, these modes can be understood as acting with a raising Virasoro generator which affects the energy of the state, however, without modifying the temperature associated with the state. Therefore only looking at the temperature of the final state it is not possible to distinguish between a black hole state and a black hole state dressed with boundary graviton modes. The conformal transformation we used in our first approach has the holographic interpretation of a (monodromy preserving) large diffeomorphism that precisely chops off this hair to yield a pure black hole as the final state of collapse. 

In this work, we present a numerical scheme to show that for generic initial state there can indeed be a large difference in energy between the initial energy injected to the system and the energy of a black hole final state estimated using thermodynamics of the black hole, with the difference in energy interpreted as the amount of energy stored in boundary gravitons.

Recently there has been a surge of work in understanding the role of soft gravitational hair in the context of black hole information paradox for flat space black holes. Interested readers can look at \cite{Pasterski:2015tva,Hawking:2016msc,Hawking:2016sgy,Strominger:2014pwa}. As we discussed, in AdS$_{3}$, already at the semiclassical level, these boundary gravitons play an important role in the late time thermalization. It is therefore natural to expect that beyond this limit, the boundary graviton modes might also play some role in understanding the black hole information paradox in the present set up. We postpone a more detailed investigation in this direction for future work.

\vskip10pt

\noindent 
The paper is organized as follows. In section \ref{sec2} we introduce our set up and the conformal block techniques which form an immensely useful tool that greatly simplifies the problem in the limit of large central charge. 

The first part of section \ref{sec3} is devoted to the understanding of the analytic continuation of the correlator to Lorentzian signature. We present here the form of the analytically continued correlator in the limit of late Lorenzian time. We also discuss in this section how to associate the behavior of the correlator at late Lorentzian time to the monodromy matrix using a gauge connection. By introducing monodromy preserving diffeomorphisms in section \ref{sec3.2} we establish the precise relation between the monodromy eigenvalues and the temperature of the final state black hole.  

In section \ref{sec4}, we present another proof of this relation using the Chern-Simons interpretation of pure gravity in AdS$_3$. 

Section \ref{numerics} contains a numerical analysis for our problem for different operator distributions. We estimate the energy of the final black holes and notice a significant deficit in energy as compared to the initial injected energy. We identify this deficit  as the fraction stored in boundary gravitons. In this numerical set up we distinguish between polyhedral and non-polyhedral distributions of operators on the  circle and discuss the consequences in either case. 

This paper has three appendices. Appendix \ref{appendix} gives a derivation of Fuchs equation starting from the Schwarzian equation for obtaining uniformizing coordinates. In appendix \ref{energyappendix} we make an estimate for the energy of the state created by insertion of finite number of heavy operators. This estimate shows that one can overcome the mass threshold necessary to obtain a stationary black hole as the final state of Lorentzian time evolution of our system of interest. Appendix \ref{mathappendix} consists of a brief discussion on how the monodromy problem of interest is closely related to the uniformization problem of a particular Riemann surface with $N$ punctures.

\section{A lightning review of  the uniformization problem} \label{sec2}
In this section we shall set up our problem and introduce the computational tools we shall be using in the rest of the paper. First we shall present the configuration of a finite number of heavy operators that represents an arbitrarily heavy state. Each of these heavy operator insertions is dual to adding a conical defect or a black hole in the bulk AdS$_{3}$. We are interested in studying the final collapse states achieved through collisions among those defects or black holes at late Lorentzian time. This collapse is captured by measuring certain correlators of probe operators in the heavy background. As we shall review, the conformal block decomposition techniques in CFT$_2$ simplifies computations of these correlation functions greatly. After presenting these techniques, the rest of the section will be devoted to presenting the main technical obstacle that we would like to overcome in this paper, namely the ``uniformization problem'' in our set up.

\begin{figure*}[h!]
    \centering
    \begin{subfigure}[b]{0.5\textwidth}
        \centering
        \includegraphics[scale=0.05]{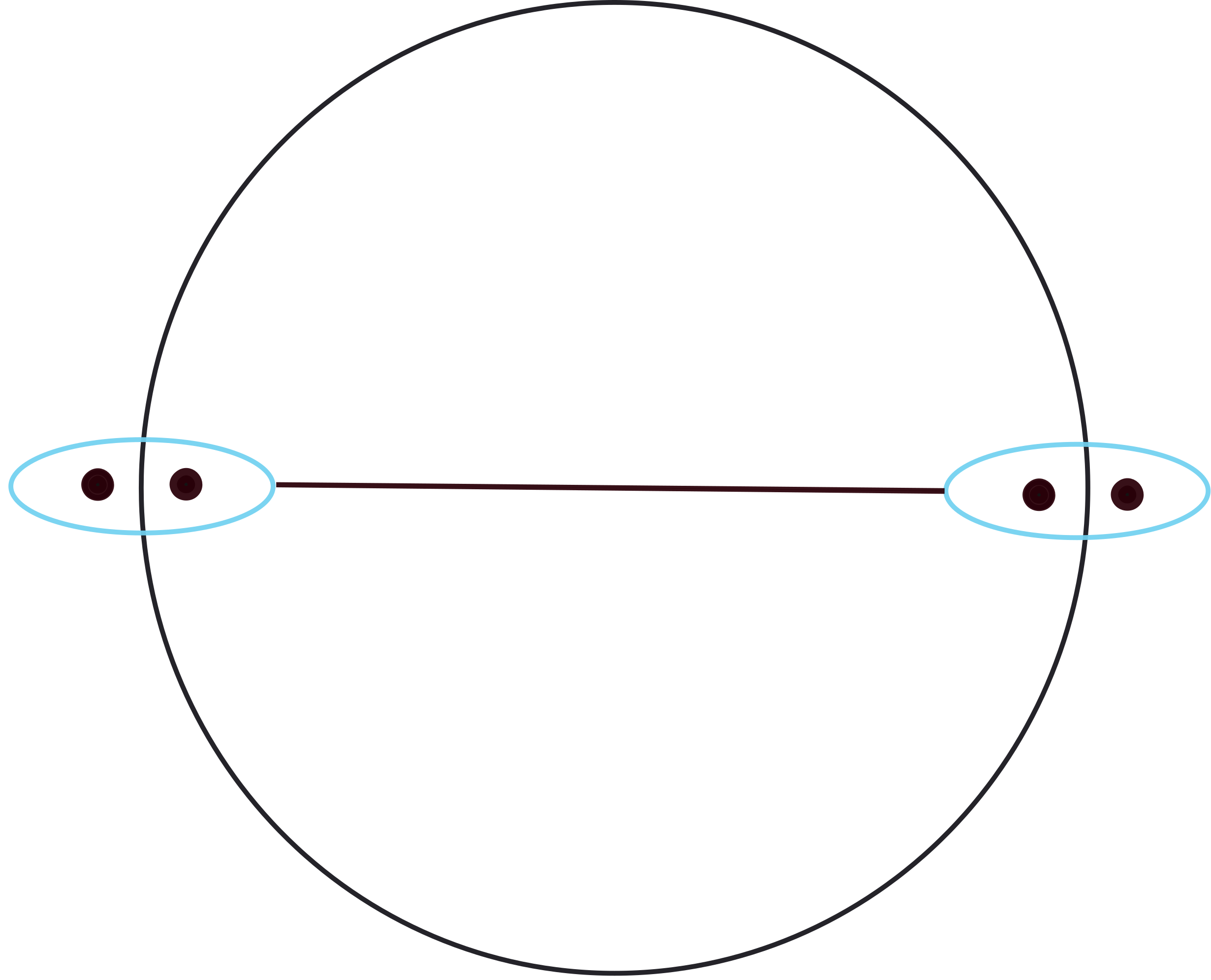}
        \caption{4 punctures, corresponding to the Heun equation}
    \end{subfigure}%
    ~ 
    \begin{subfigure}[b]{0.5\textwidth}
        \centering
        \includegraphics[scale=0.05]{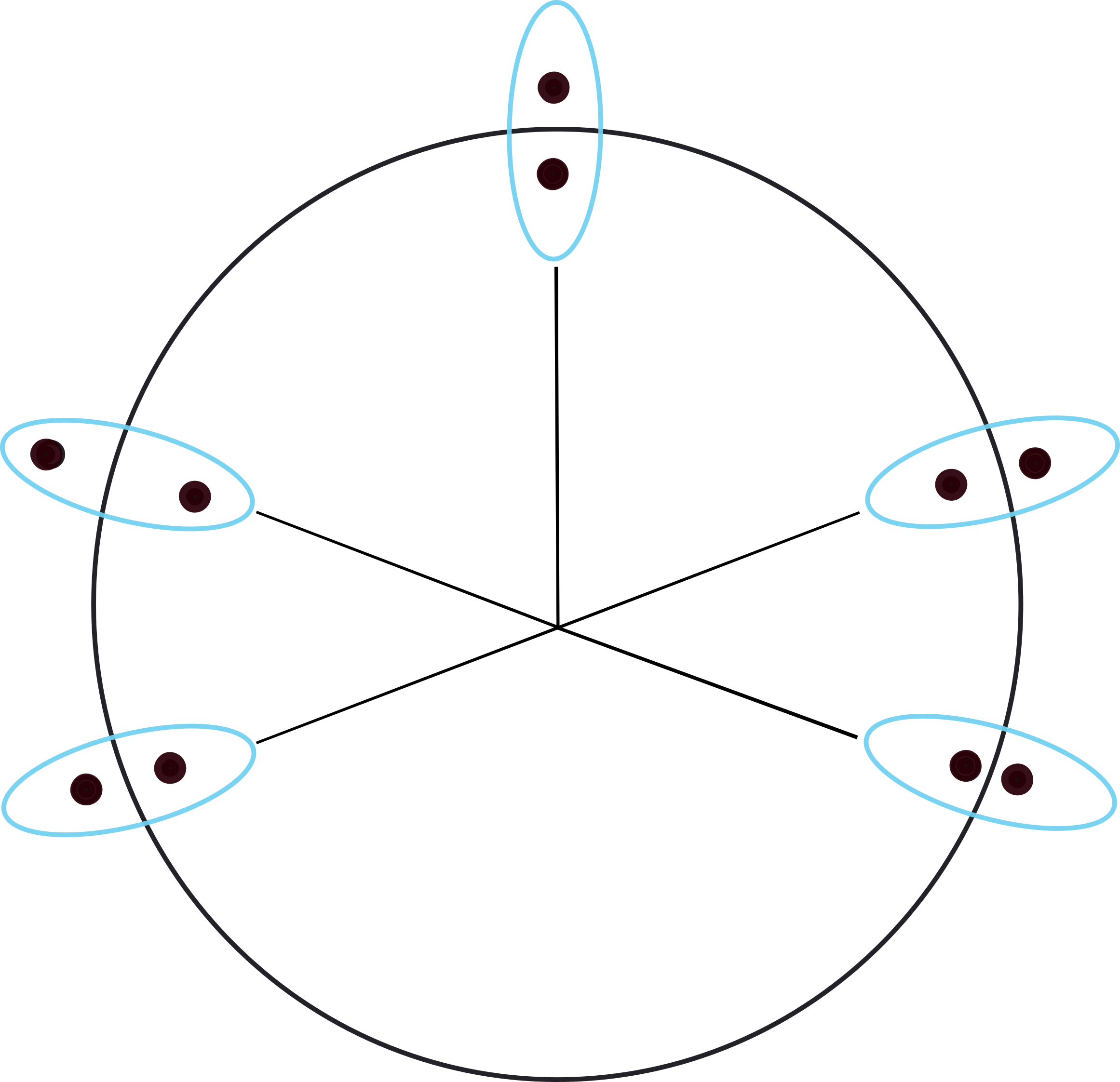}
        \caption{surface corresponding to a Riemann surface with 10 punctures}
    \end{subfigure}
		\caption{This diagram displays the type of mirror pair configurations under consideration, and displays the OPE channel used. The dots represent heavy operator insertions on the radial plane. The left configuration potentially forms a stationary black hole, the right configuration is expected to not be stationary at late time. }
   \label{pictures1}
\end{figure*}

\subsection{Basic set up}

We consider a generic heavy state $|V\rangle$ created by acting with heavy primary operators, $O_H$ of scaling dimensions $H=\mathcal{O}(c)$ at the same time but at different points in space. In radial quantization,
\be 
\label{V-state}
|V\rangle=O(x_1)...O(x_n)|0\rangle, \hspace{15mm} |x_n|=1-\sigma_n.
\ee
Here $x_i$'s are complex coordinates in Euclidean CFT$_2$ so that the CFT states lie on a unit circle. To avoid UV divergences, we regularize our operator insertions $x_n$ by shifting them a small distance $\sigma$ away from the unit circle ($|x_n|=1-\sigma_n$) \cite{Anous:2016kss}. The respective adjoint operators contained in the bra-state $O^{\dagger}(1/\bar{x}_n)$ are therefore located on the circle with radius $(1-\sigma_n)^{-1}$. We call a heavy primary insertion and its adjoint together, a ``mirror pair''. We shall frequently use this nomenclature later in this work. We consider arbitrary distribution of these mirror pairs on the circle. Configuration of these mirror pairs of operators in a regular polyhedral distributions on the circle is a sufficient condition that the resulting black hole after collision will be at rest with respect to the centre of AdS while relaxing the same yields oscillating black holes in the final state of evolution. Both these configurations are shown in figure \ref{pictures1}. Using numerical techniques, later in this work, we will show the physical consequences of and difference between having a static and a non-static configurations of operators explicitly.

As we mentioned before, we want to study the propagation of small probes on this heavy background that can be interpreted as undergoing gravitational collapse to a black hole state. To achieve we will be interested in the qualitative properties of correlators of the form  \eqref{correlatorN}. In particular, we shall focus on 2-point correlation function 
\begin{equation}
\mathcal{A}=\langle V|Q(z_1)Q(z_2)|V\rangle,
\label{correlator11}
\end{equation}
where $Q(z_i)$'s are light operators with scaling dimensions $h_Q\ll c$ and do not backreact on the background heavy state. 

At this point it is worth reminding ourselves that, to reach the collapse state, we need to study evolution of the correlation function, \eqref{correlator11} to late Lorentzian time. This requires analytic continuations of \eqref{correlator11} to different Lorentzian correlators corresponding to different operator orderings. As we shall discuss in next section, for our purpose we will specifically need out-of-time-ordered correlation functions. 

The leading semi-classical information about the final collapse state is encoded in the regime of large central charge. In order to extract this information, it is convenient to use the monodromy method developed in \cite{Hartman:2013mia} and used in \cite{Anous:2016kss} in a related context. However, it is worth mentioning here once again that the continuous state limit considered in \cite{Anous:2016kss} stands as a very special case of the generic scenario we consider in this work. As discussed in the introduction, the large number of operator insertion, $n \rightarrow \infty$, as opposed to the finite number of heavy operator insertions we consider in \eqref{V-state}, drastically simplifies the computation. The ``heaviness'' of inserted operators in our case naively destroys the perturbative expansion at large $c$ . In the next subsections we shall review this problem and an effective solution thereof prescribed in a set of papers, \cite{Fitzpatrick:2015zha, Fitzpatrick:2015dlt}.
However, as mentioned earlier, we allow an arbitrary initial energy distribution corresponding to insertions of any finite number of heavy operators. We will, therefore, find that the afore-mentioned prescription does not completely fix the semi-classical correlator for us, although it simplifies our problem to a large extent.

\subsection{The vacuum block and semiclassical correlation functions}
In a 2d CFT with large central charge and sparse spectrum of low-lying operators, the method of conformal block decomposition provides the most efficient way to study the correlator given in \eqref{correlator11}. 
In general, in any quantum field theory, a correlator can be rewritten by inserting a projection operator made out of a complete set of states as 
\begin{equation} \nonumber
\begin{aligned}
\mathcal{A}(x_1,...,x_n) &\equiv \langle O(x_1)...O(x_n)O(\bar{x}_1^{-1})...O(\bar{x}_n^{-1})Q(z_1)Q_(z_2)\rangle \\ &= \sum_{\alpha} \langle O(x_1)...O(x_n)O(\bar{x}_1^{-1})...O(\bar{x}_n^{-1})|\alpha\rangle \langle \alpha | Q(z_1)Q(z_2)\rangle
\end{aligned}
\end{equation}
In a 2d CFT, in particular, one can arrange the states into irreducible representations of the conformal algebra, namely the Virasoro algebra. We can formally decompose the projector into a sum over partial projectors associated with each irreducible representation contained in the CFT spectrum. The full conformal algebra in $d =2$ consists of two copies of Virasoro algebra. In what follows, we will present formulae only for the holomorphic sector, however, keeping in mind that all these statements will have anti-holomorphic counterparts. 
\begin{equation} 
P=\sum_{h} P_{h}= \sum_{h} \sum_{\{n_i,k_i\}} \frac{L_{-m_n}^{k_n}...L_{-1}^{k_1}|h\rangle\langle h|L_1^{k_1}...L_{m_n}^{k_n}}{\langle h|L_1^{k_1}...L_{m_n}^{k_n}L_{-m_n}^{k_n}...L_{-1}^{k_1}|h\rangle},
\label{projector}
\end{equation}
where we have used the fact that a single Verma module of the conformal algebra can be generated by acting on the primary state with all ordered combinations of raising Virasoro generators
$$
\mathcal{V}_h=\{ L_{-n}^{k_n}....L_{-1}^{k_1}|h\rangle \}.
$$
The denominator in \eqref{projector} is the normalization factor.
In order to avoid an over-complete basis, one needs to adopt an ordering convention. The conventional choice is given by $m_1 > m_2 > \cdots > m_n$.
The modes, $L_n$ are subject to the usual Virasoro algebra
\be
\label{con-alg}
[ L_n, L_m ] = (n-m)L_{n+m} + \frac{c}{12} n(n^2-1)\delta_{n,-m}.
\ee
The projection of a correlator on one particular irreducible representation of the conformal algebra constitutes a single conformal block. Schematically, for the holomorphic part 
\be
\label{conblo1}
\langle O(x_1)...O(x_n)O(\bar{x}_1^{-1})...O(\bar{x}_n^{-1}) P_h Q(z_1)Q(z_2)\rangle = {\mathcal{C}} \alpha_h  {\mathcal{F}}_h,
\ee
where $ {\mathcal{C}}$ denotes a choice of normalization and $\alpha_h$, a kinematic factor given by a product of OPE coefficients in a particular OPE channel of interest.  ${\mathcal{F}}_h$ is the holomorphic conformal block which, on the other hand, is entirely fixed by the Virasoro symmetry. A similar expression also holds for the anti-holomorphic part.
Each of the representations of the Virasoro algebra is associated with a primary operator, ${\cal O}_{h,{\bar h}}$ that creates the state $|h\rangle$ acting on the vacuum state,
\be
\label{op-sep}
{\cal O}_{h,{\bar h}} |0\rangle = |h\rangle \otimes |{\bar h}\rangle.
\ee 
Substituting the holomorphic part of \eqref{op-sep} in \eqref{conblo1}, one can express the conformal block in terms of factorized correlators in the vacuum.

In general the computation of conformal blocks in 2d CFT simplifies greatly in the regime where the central charge becomes very large, i.e the semi-classical regime. The simplification most clearly appears when the representation of the Virasoro block contains operators, $O(x_i)$ and intermediate operators, ${\cal O}_{h,{\bar h}}$ whose scaling dimensions $h, \bar{h}$ are small with respect to the central charge $c$. In this case, while the numerator of \eqref{projector} does not scale with $c$, from the conformal algebra, $\eqref{con-alg}$, the denominator provides factors of $c$ when the generators are not contained within the global conformal group ($\{ L_{-1},L_0,L_1\}$). In the large $c$ limit, the leading semi-classical result can therefore be extracted solely from the global conformal block. Contribution from higher descendents will be suppressed by factors of $c$. 

However, as mentioned before, we shall be considering heavy states created by a finite number of heavy operators, i.e. operators whose scaling dimension is proportional to the central charge $c$. As a result the numerator factors of \eqref{projector} will contain operators with scaling dimensions scaling with $c$. However, if we allow this, the numerator of \eqref{projector} will produce factors scaling with powers of $c$ that can overcome the suppression from the denominator. This can be demonstrated by noting that the Virasoro generators are defined as the coefficients of the Laurent series of the stress-energy tensor ${\mathcal T}(z)$.
$$
{\mathcal T}(z)=\sum_{n=-\infty}^{\infty}z^{-n-2}L_n, \hspace{15mm} L_n=\frac{1}{2\pi} i \oint dz\, z^{n+1}{\mathcal T}(z)
$$
We can therefore construct a generating function for correlators containing Virasoro generators
\begin{multline}
\langle O(x_1)...O(x_n)O(\bar{x}_1^{-1})...O(\bar{x}_n^{-1})L_{-n_m}...L_{-n_1} {\cal O}_h(0) \rangle\\= \oint dz_1...dz_m\, z_1^{n_1+1}...z_m^{n_m+1}\langle O(x_1)...O(x_n)O(\bar{x}_1^{-1})...O(\bar{x}_n^{-1}) {\cal O}_h(0){\mathcal T}(z_1)...{\mathcal T}(z_m)\rangle.
\label{corr}
\end{multline}
By means of the Virasoro Ward identity we can see that the right-hand side will contain terms proportional to powers of the scaling dimensions of operators which in turn scale as powers in $c$. 

Under the circumstances, it becomes impossible to extract the leading behavior from the global conformal block as the perturbative suppression of contributions coming from higher descendent states no longer works.


In \cite{Fitzpatrick:2015zha} and \cite{Fitzpatrick:2015dlt}, the authors came up with an idea to circumvent this problem. They established that the lack of suppression of the terms containing higher numbers of Virasoro modes is not a fundamental problem but rather signifies a poor choice of conformal frame. The authors exploited the fact that any metric in a $2$-dimensional spacetime is conformally related to the flat space. The proposal was to choose an appropriate geometry, or equivalently employ a conformal transformation on the complex coordinates, $z \rightarrow w(z)$  in such a way that the correlator in the new ``$w$''-coordinate system, 
\be
\label{wfn}
\langle O(x_1)...O(x_n)O(\bar{x}_1^{-1}) \cdots O(\bar{x}_n^{-1})  {\cal O}_h(w){\mathcal T}(w_1)...{\mathcal T}(w_m)\rangle
\ee  
does not scale with positive powers of $c$.  This can be done by exploiting the inhomogenous transformation of the stress energy tensor under conformal transformation, namely
\be
\label{Tw1}
{\mathcal T}(w)=\left(\frac{dz}{dw}\right)^{2}{\mathcal T}(z(w))+\frac{c}{12}S[z(w),w],
\ee
where $S[z(w),w]$ is the Schwarzian derivative defined by
\be
\label{schwarzian}
S[z(w),w]\equiv\frac{z'''(w)}{z'(w)}-\frac{3}{2}\left(\frac{z''(w)}{z'(w)}\right)^2.
\ee
The Schwarzian extension of the transformation \eqref{Tw1} ensures that in the new coordinate system the projector states constructed out of the global conformal block give O(1) contributions, while  contributions coming from the states outside of the global block remain suppressed by factors of 1/c. Furthermore, it can be proved using the ${\mathcal T}(w1){\mathcal T}(w2)$ OPE that if the correlator with a single stress energy tensor insertion, $\langle O(x_1)...O(x_n)O(\bar{x}_1^{-1}) \cdots O(\bar{x}_n^{-1})  {\cal O}_h(w){\mathcal T}(w)\rangle$ can be expressed as a systematic expansion in $1\over c$, similar expansion for an arbitrary number of stress energy tensor insertions as in \eqref{wfn} is automatically guaranteed.

One simplification that we adopt in our analysis is that we focus on the vacuum block corresponding $h=0 , {\bar h} = 0$ operators, namely the identity operator. This is reflected in the OPE channel depicted in figure 1 where all operators are fused with their adjoint and the resulting identity operator are in turn fused as indicated by the lines. From the gravitational bulk perspective of the AdS$_{3}$/ CFT$_{2}$ correspondence, all graviton modes are expected to correspond to descendant states of the vacuum. Thus all exchanged gravitational modes are encompassed within the exchange of this one particular conformal block. From the perspective of CFT this is a consequence of large $c$ and sparse spectrum of low-lying single trace operators. For certain conformal field theories this can be explicitly proved \cite{Hartman:2013mia, Hartman:2014oaa}. We will assume this to work for generic CFTs having holographic duals since the afore mentioned properties typically hold in those CFTs \cite{Heemskerk:2009pn, Fitzpatrick:2010zm, ElShowk:2011ag}.

In this case the irreducible representation of the global conformal group contains the vacuum state only. It would therefore be sufficient to choose a conformal transformation that renders correlators with single stress-energy tensor insertion vanish identically, namely,
\begin{equation}
\langle O(w(x_1))...O(w(x_n))O(w(\bar{x}_1^{-1}))...O(w(\bar{x}_n^{-1})){\mathcal T}(w)\rangle=0.
\label{uniformizingproperty}
\end{equation} 
If this property holds for the coordinate system $w(z)$ this will ensure that it is only the contributions to the partial projector from the global conformal group that survive the limit $c\rightarrow \infty$. As a result, in the new $w$-coordinate system, the vacuum conformal block corresponds to the factorized correlator
\begin{equation}\nonumber
\begin{aligned}
A_0(w_1,w_2)&=\langle O(w(x_1))...O(w(x_n))O(w(\bar{x}_1^{-1}))...O(w(\bar{x}_n^{-1}))\, P_{0}\, Q(w_1)Q(w_2)\rangle\\&=\langle O(w(x_1))...O(w(x_n))O(w(\bar{x}_1^{-1}))...O(w(\bar{x}_n^{-1}))\rangle\,\langle Q(w_1)Q(w_2)\rangle.
\end{aligned}
\end{equation}
The correlation function factor containing the heavy operators does not contain any dependance on $w_i$ so it is an overall factor that we will henceforth ignore. We can transform back to the original $z$-coordinates by means of the usual transformation rule of primary operators
\begin{equation}
A_0(z_1,z_2)=\left(\frac{dw}{dz_1}\right)^{h_{Q}}\left(\frac{dw}{dz_2}\right)^{h_{Q}}\left(w(z_1)-w(z_2)\right)^{-2h_{Q}},
\label{vacuumblock}
\end{equation}
this will be the object that will be studied for the remainder of this paper.

\subsection{The defining equation for the uniformizing coordinates}
In this subsection we will focus on how to construct these uniformizing coordinates. It is known that the stress-energy tensor satisfies the following transformation law given by \eqref{Tw1} and \eqref{schwarzian}. 

Let us define the ``stress-energy function''
$$
T(z)=\frac{\langle O(x_1)...O(x_n)O(\bar{x}_1^{-1})...O(\bar{x}_n^{-1}){\mathcal T}(z)\rangle}{\langle O(x_1)...O(x_n)O(\bar{x}_1^{-1})...O(\bar{x}_n^{-1})\rangle}
$$
The defining property of the uniformizing coordinates \eqref{uniformizingproperty} can be transformed back to the original radial plane coordinates $z$ by means of the transformation law
$$
T(z) z'(w)^2+\frac{c}{12}\left(\frac{z'''(w)}{z'(w)}-\frac{3}{2}\left(\frac{z''(w)}{z'(w)}\right)^2\right)=0.
$$
It is explicitly demonstrated in the appendix \ref{appendix} that the resulting Schwartzian differential equation for the uniformzing geometry can be inverted and rewritten into a second order linear differential equation of Fuchs class.
\begin{equation}
u''(z)+\frac{6}{c}{T}(z)u(z)=0,
\label{fuchs}
\end{equation}
where $u(z)$ is given by $u(z)^{-2}=dw/dz$. Typically $T(z)$ is a meromorphic function whose poles are regular singular points of the differential equation, but we will consider as an example the case of \cite{Anous:2016kss} where this analyticity condition is relaxed. In terms of $u(z)$ the vacuum conformal block \eqref{vacuumblock} can be written as
\begin{equation}
A_0(z_1,z_2)=u(z_1)^{-2h_{Q}}u(z_2)^{-2h_{Q}}\left( \int_{z_1}^{z_2} u(z)^{-2} dz\right)^{-2h_{Q}}.
\label{correlator-uni}
\end{equation}
In the next section we will demonstrate that a lot of qualitative properties can be derived from the form of this formula without the need to actually explicitly solve the Fuchs type equation \eqref{fuchs} above.

\section{Lorentzian time-evolution on the radial plane}\label{sec3}
The eventual goal of this paper is to describe the dynamics of probe correlators at asymptotically late Lorentzian times. This matter is complicated by the fact that the probe correlators
$$
\langle V(0)|Q(t_1)Q(t_2)|V(0)\rangle=\langle V|e^{-iHt_1}Qe^{-iH(t_2-t_1)}Qe^{-iHt_2}|V\rangle
$$
are out of time order\footnote{For notational convenience we gave the state $|V(0)\rangle$ a single time label. As mentioned in the previous section we do not want to restrict ourselves to heavy operators inserted at the same time, therefore the appropriate way to think of $t=0$ is as the time of the latest heavy operator inserted and the effect of all other heavy operators on the state $|V(0)\rangle$ is included through a path integral.}. A consequence of the reconstruction theorems of quantum field theory \cite{Osterwalder:1974tc, Luscher:1974ez} states that all operator orderings within a correlator are related to each other through analytic continuation \cite{Streater:1989vi}. Hence the problem is to evaluate the full analytically continued correlator on the correct branch. To distinguish the correct branch we give the operator locations small imaginary parts corresponding to their ordering within the correlator. The Heisenberg picture above elucidates the signs and relative magnitudes of the imaginary parts as there exists a unique choice of either upper or lower half plane to which the time differences can be extended in order to obtain a convergent correlator on the extended domain.  In this case we will give the Lorentzian times small imaginary parts, which can equivalently be thought of as giving them small real shifts in Euclidean time instead. This suggests the following algorithm\footnote{ See \cite{Roberts:2014ifa} for an explicit demonstration of Lorentzian continuation of Euclidean correlators in a similar context.}
\begin{itemize}
\item Give all Lorentzian times imaginary parts such that the time differences fall within the appropriate extended domain.
\item Set all real parts of the Lorentzian times equal to zero, the resulting correlator is purely Euclidean and can be computed using standard CFT techniques.
\item As a final step bring all real parts of Lorentzian time (or equivalently imaginary parts of Euclidean time) back to their original values. The imaginary parts will prescribe how to circumnavigate all encountered branch points and the associated multi-valuedness of the Lorentzian correlator.
\end{itemize}
This method provides our strategy to deriving results at late Lorentzian time. As mentioned before, we evaluate our correlator in radial quantization. We make use of the exponential map to see how our shifts in Euclidean time manifest on the radial plane.
$$
z=e^{\tau+i\phi}, \hspace{1cm} \bar{z}=e^{\tau-i\phi}. 
$$
One can see that a small shift in Euclidean time corresponds to a small radial shift on the radial plane. This justifies the $\sigma_i$ regularization scheme presented in the last section. If the heavy state operators and their adjoints are respectively shifted radially away from or toward the origin while maintaining the position of the probe operators on the unit circle we ensure that we are on the right branch after reinstating the appropriate Lorentzian times. Hence from then onwards Lorentzian time-evolution of the probe correlators simply corresponds to letting them perform cycles along the unit circle.

\begin{figure}
	\centering
		\includegraphics[scale=0.5]{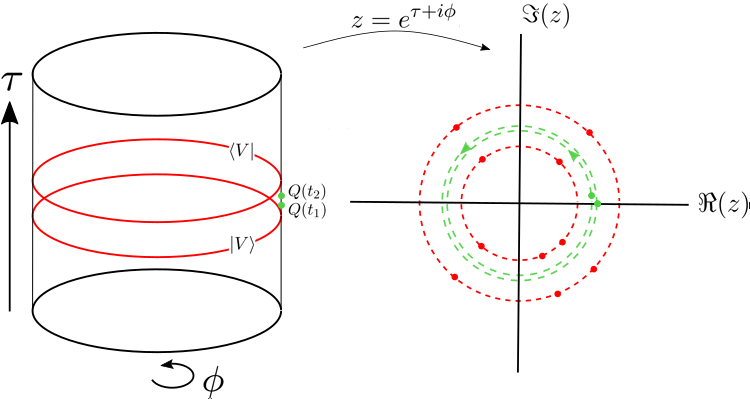}
	\caption{This diagram illustrates the effect of the exponential map on states created on the Euclidean cylinder. The red dots correspond to the insertions of heavy operators while the green dots denote light probe operators.}
	\label{fig:ExponentialMap}
\end{figure}

\begin{figure}
	\centering
		\includegraphics[scale=0.5]{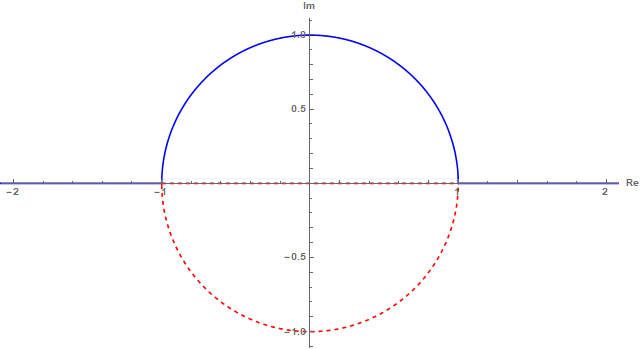}
	\caption{The restriction of the unit circle monodromy matrix to $SU(1,1)$ restricts the eigenvalues it can possess to be either purely real or pure phases. Since the eigenvalues are each others inverse $\lambda_1=1/\lambda_2$, the domain in which the eigenvalues can reside can be further split up into two independent domains indicated by the blue solid line and red dashed line.}
	\label{EigenvalueDomain}
\end{figure}

This implies that this cycle needs to possess a non-trivial monodromy in the case where the correlation function is not expected to be periodic in time. Let us return to the Fuchsian equation \eqref{fuchs}, it is a second-order linear differential equation which means that it's general solution is a linear combination of two particular solutions
$$
u(z)=c_1 u_1(z)+c_2 u_2(z).
$$
Since $T(z)$ is not analytic everywhere on the unit disk the linear space spanned by the particular solutions possesses a non-trivial flat connection on the unit circle. This statement can be made visible by rewriting the single second-order ODE \eqref{fuchs} into two coupled first-order ODEs\cite{Castro:2013lba}
$$
\partial_{z}U(z)=a(z)U(z),
$$
here $U(z)$ is the fundamental matrix of solutions and $a(z)$ is the connection matrix
\begin{equation}
U(z)=\begin{pmatrix}
u_1(z)&u_2(z)\\
\end{pmatrix}, \hspace{12mm}
a(z)=\begin{pmatrix}
0&1\\
-\frac{6}{c}T(z)&0\\
\end{pmatrix}.
\label{connection}
\end{equation}
In principle the general solution of this set of ODEs can be written as an initial value problem
$$
U(z)=\mathcal{P}\left\{e^{\int^{z} dz'\, a(z')}\right\}U(0),
$$
here $\mathcal{P}$ designates the path-ordered integral. Hence the monodromy relevant for Lorentzian time-evolution is given by
\begin{equation}
M\sim\mathcal{P}\left\{e^{\oint_{|z|=1} dz\, a(z)}\right\},
\label{pathorderedintegral}
\end{equation}
here $\sim$ indicates equal up to a similarity transformation. From the fact that $a(z)$ is always traceless we can establish the general property that the monodromy matrix $M$ possesses unit determinant, i.e. as a matrix $M$ in contained within the group $SL(2,\mathcal{C})$. In principle one can go further, the symmetric distribution of heavy operators and their adjoints around unit circle provides the following reflection property
\begin{equation} \nonumber
T(z)=z^{-4}\overline{T(1/\bar{z})}.
\end{equation}
As rather elegantly proved in \cite{Hulik:2016ifr}, this property implies that in general the monodromy matrix $M$ is up to a similarity transformation contained within the group $SU(1,1)$. Since similarity transformation do not effect the eigenvalues this restricts the eigenvalues of $M$ to be either pure phase or purely real (see figure \ref{EigenvalueDomain}). This divides $M$ into one of three classes purely real eigenvalues, purely imaginary eigenvalues and two cross-over point at 1 and -1 where the eigenvalues are degenerate.

The path-ordered integral over the flat connection above has a natural interpretation in the Chern-Simons formulation of gravity in $AdS_3$. We will exploit this interpretation in section 4.

\subsection{Late-time behavior of the correlators}
It is interesting to see what kind of algebraic effect the considerations of the previous section have on \eqref{correlator11} in the late-time limit which is obtained by letting the probe correlator undergo a large number of cycles on the unit circle. Let us assume a basis of solutions that diagonalizes the monodromy matrix around the unit circle\footnote{This can be done whenever the eigenvalues are non-degenerate, which are our cases of interest}. Since the monodromy has to have unit determinant, it only has one free eigenvalue $\lambda$ the other one has to be its inverse $1/\lambda$. After $n$ cycles the general solution to the Fuchs equation will generically be given by
$$
u\left(e^{2\pi i n}z\right)=\lambda^n c_1 u_1(z)+\lambda^{-n} c_2 u_2(z),
$$
we here assume, without loss of generality, that $|\lambda|\geq|1/\lambda|$. There are two particular cases of interest. Either both eigenvalues are distinct and pure phase, in which case time-evolution in Lorentzian time will yield periodic correlation functions. The physical interpretation is that the system fails to collapse into a black hole and the heavy material will continue to oscillate forever around the centre of AdS$\,_3$. The monodromy falls into elliptic class by the classification in \cite{Martinec:1998wm}. 

The second case is when $|\lambda|> 1$, which is most interesting scenario for us. Since $n$ plays the role of a discretized time, the $u_1(z)$ mode acts as a stable mode, whereas $u_2(z)$ plays the role of a decaying mode. We find that the mode $u_1(z)$ should be given by a solution that corresponds to a BTZ black hole, since this is the situation that the collapse process should converge to at late times. Furthermore if both $z_1$ and $z_2$ make the same number of cycles the constant $\lambda$ drops out of the correlator \eqref{correlator11}, corroborating the interpretation as a stable mode.

The eigenvalue is not purely a mathematical construct, it can be given a physical interpretation by considering the late-time limit. Consider $z_1=e^{2\pi i n}z_0,;\; z_2=e^{2\pi i (n+m)}z_0$, where $z_0$ is some number on the unit circle on the principle sheet and $n\gg 1$. By allowing $z_2$ to make a few more cycles compared to $z_1$, corresponding to time-advancing $Q(z_2)$ we can obtain the following result for the correlator
\begin{multline} \nonumber
\langle O(x_1)...O(x_n)Q(z_1)Q(z_2(n))\rangle = u(z_1)^{-2h_Q}\lambda^{-2mh_Q}u(z_2)^{-2h_Q}\\\times\left(\int_{z_1}^{e^{2\pi i}z_1}u(z)^{-2}\, dz + \int_{e^{2\pi i}z_1}^{e^{4\pi i}z_1}\lambda^{-2} u(z)^{-2}\, dz + ...+\int_{e^{2\pi i(m-1)}z_1}^{e^{2\pi i m}z_1}\lambda^{-2(m-1)}u(z)^{-2}\, dz\right)^{-2h_Q}\\ = u(z_1)^{-2h_Q}u(z_2)^{-2h_Q}\lambda^{-2mh_Q}\left(\frac{\lambda^{-2m}-1}{\lambda^{-2}-1}\right)^{-2h_Q}
\left( \int_{z_1}^{e^{2\pi i}z_1}u(z)^{-2}\, dz \right)^{-2h_Q}.
\end{multline}
Hence as $m$ becomes larger we note the approximate scaling law
$$
\langle O(x_1)...O(x_n)Q(z_1)Q(z_2(m))\rangle \sim \lambda^{-2mh_Q}.
$$
It is worth noting that the scaling behaviour is independent of $n$ and only depends on the difference between the number of cycles which can be interpreted as the difference in time when the two operators are fixed at the same spatial point. Such a scaling relation at late times is suggestive of thermal behavior. In the upcoming sections we will support this observation by two separate arguments. Firstly we will demonstrate that when the eigenvalues are purely real we can construct a monodromy preserving diffeomorphism to a stress-energy tensor whose associated Fuchs equation can be solved. 

Secondly we will demonstrate that the eigenvalue $\lambda$ has a natural manifestation in the Chern-Simons formulation of $3d$ Einstein gravity, as being related to the temperature of the bulk black hole geometry associated to the boundary connection $a=\left(J_{-1}+\frac{6}{c}T(z)J_{1}\right)dz$, where the $J_i$ matrices are generators of $SL(2,\mathbb{R})$. Either way we will find out that the explicit relationship is given by
\begin{equation}
\log(|\lambda|)=2\pi^2 T_{BH},
\label{scalinglaw}
\end{equation}
with the black hole temperature $T_{BH}$. If we equate $2\pi m$ with time difference $(t_2 - t_1)$ we find the following asymptotic scaling
\begin{equation}
A_0(t_1,t_2)\sim e^{-2\pi h_Q T_{BH} (t_2-t_1)},
\label{bhcorrelator}
\end{equation}
where we have put the two operators at the same spatial point.
This is exactly the form of a thermal correlator with temperature $T_{BH}$. This is our main conclusion, without solving the Fuchs equation we find that at asymptotic late Lorentzian time the probe correlator converges to a thermal correlator with a temperature given by the temperature of the bulk theory black hole.

\subsection{Monodromy preserving diffeomorphisms and their bulk interpretation as boundary gravitons} \label{sec3.2}
In this section we provide the first proof that the hyperbolic class of asymptotic Lorentzian correlators correspond to thermal black hole correlators. The proof provided here is the most straightforward of the two, but also the most abstract. As a consequence it clarifies that the bulk final state consists of a black hole dressed by boundary gravitons. Here we will exploit the fact that the stress-energy functions, $T(z)$ for which the associated Fuchs equations, \eqref{fuchs} possess monodromy matrices around the unit circle with real eigenvalues must to be contained within a specific Virasoro coadjoint orbit  \cite{Witten:1987ty}. Different points in the orbit are related by conformal transformations. In the dual gravity picture, this orbit can be associated with a class of black hole solutions to the Einstein equations.

It is well-known that any asymptotically AdS$_3$ solution in the Fefferman-Graham gauge is given by the Banados geometries \cite{Banados:1998gg}
\begin{equation}\nonumber
ds^2=\frac{dr^2}{r^2}-\left(rdz-\frac{\bar{L}(\bar{z})}{r}d\bar{z}\right)\left(rd\bar{z}-\frac{L(z)}{r}dz\right),
\end{equation}
where $L(z)$ and $\bar{L}{(\bar z)}$ are single-valued functions related to the holomorphic and anti-holomorphic stress tensors of the boundary CFT. After fixing the Banados gauge there exists a residual symmetry that preserves the general form of this metric but changes the functions $L(z)$ and $\bar{L}(\bar{z})$. At $r\rightarrow\infty$ the vector fields generating these coordinate transformations are given by \cite{Brown:1986nw}
\begin{equation} \nonumber
\zeta=\left(\xi(z)+\frac{1}{2r^2}\partial_{\bar{z}}^2\bar{\xi}(\bar{z})\right)\partial_{z} \, +\, \left(\bar{\xi}(\bar{z})+\frac{1}{2r^2}\partial_{z}^2\xi(z)\right)\partial_{\bar{z}} \, - \, \frac{r}{2}\left(\partial_{z}\xi(z)+\partial_{\bar{z}}\bar{\xi}(\bar{z})\right)\partial_{r}.
\end{equation}  
The set of all these vector fields is parametrized by the two functions $\xi(z)$ and $\bar{\xi}(\bar{z})$. It can be shown that under the infinitesimal coordinate transformations generated by these vector fields the functions $L(z)$, $\bar{L}(\bar{z})$ transform according to the infinitesimal Virasoro Ward identity
\begin{equation} \nonumber
\delta_{\zeta}L(z)=2L(z)\partial_{z}\xi(z)+\xi(z)\partial_z L(z) -\frac{1}{2}\partial_{z}^3\xi(z),
\end{equation}
with a similar rule holding for $\bar{L}(\bar{z})$. From this it is natural to identify the functions parametrizing the Banados metric with the boundary stress tensor through $L(z)=\frac{6}{c}T(z)$. From this one can conclude that the dual description of (single-valued) conformal transformations is given by large diffeomorphisms\footnote{Large in the sense that they do not drop of fast enough at infinity.} that preserve the asymptotic boundary conditions of the AdS$_3$. Hence stress tensors that are connected to each other through single-valued coordinate transformations correspond in the bulk to geometries that are related to each other through boundary gravitons.

Given a function $T(z)$ (correspondingly $L(z)$) it is natural to consider the set of all other functions $T(z)$ that can be obtained by single-valued conformal transformation. To remind ourselves the infinitesimal transformation law integrates to one of the form of \eqref{Tw1}, namely, 
$$
T(w)=\left(\frac{dw}{dz}\right)^2T(z(w))+\frac{c}{12}S[z(w),w],
$$
for a finite conformal transformation. The orbits of $T(z)$ under all single-valued conformal transformations is directly related to the Virasoro coadjoint orbits discussed in the beginning of the section. The Virasoro orbits have all been classified in terms of the monodromy of the solutions Hills equation \cite{Balog:1997zz} (see table \ref{orbits}). Up to a change of variables Hill's equation is the Fuchs equation whose domain has been restricted to the unit circle. Each orbit is classified by a reference point, a function $T(z)$, usually of some convenient form, that can be reached by conformal transformation.

Of special interest are the the functions $T(z)$ that generate solutions to the Fuchs equation whose monodromy matrix has real eigenvalues, these coadjoint orbits contain a reference point that is the standard BTZ geometry \cite{Compere:2015knw, Garbarz:2014kaa, Sheikh-Jabbari:2016unm}. From the CFT perspective this reference point corresponds to a stress tensor expectation value associated to a state created by acting on the vacuum with a primary operator whose scaling dimension satisfies $H>c/12$.

\begin{table}
\begin{tabular}{l||l|l|l}
Class & eigenvalues & Coadjoint orbit & dual geometry\\
\hline 
\hline
Elliptic & pure phase $(\lambda \neq 1,-1)$& $\mathcal{C}_{\nu}$ \, $0<\nu<1$ &  Conical defect \\
\hline
Parabolic & $\lambda=1$ or -1 & $\mathcal{P}_0^{+}$, $\mathcal{P}_1^{-}$  & minimal mass BTZ \\
\hline
Hyperbolic & purely real & $\mathcal{B}_{0}(b)$ & BTZ black hole \\
\hline
Exceptional & $\lambda=1$, &  $\mathcal{E}_1$ & Vacuum \\
\end{tabular}
\caption{A characterisation of the relevant monodromy classes their associated Virasoro coadjoint orbit and the geometry of the holographic dual. (following \cite{Balog:1997zz},\cite{Martinec:1998wm})}
\label{orbits}
\end{table}

From the boundary CFT perspective it is not entirely obvious that the generic stress tensor we are considering is contained within an orbit that contains as a reference point a stress tensor associated to a state created by a single primary operator. First of all, it would naively be expected that a generic heavy CFT state would correspond to a linear combination of energy eigenstates. However, on the other hand, the existence of the BTZ reference point suggests that even a generic heavy stress energy function is dominated by contributions coming from a single conformal family. These two apparently conflicting CFT statements can  actually be justified in the large $c$ limit. In appendix \ref{energyappendix} we have provided an explicit example towards the justification. There we consider a state generated by $n$ heavy operators inserted on some radial slice (see appendix \ref{energyappendix} for more details). As the energy of our state increases the variance in the energy increases as well. However, the variance increases parametrically less fast than the expectation value ($\sqrt{nh/\sigma}$ as opposed to  $nh/\sigma$). As one can check, for a heavy state with an energy expectation at least of the order of the central charge the energy distribution can be well approximated as being sharply peaked. This is of course directly analogous to the intuition of the eigenstate thermalisation hypothesis \cite{srednicki1999approach}. Therefore the general expectation is that a pure state created by acting with the heavy primary operators on the vacuum is dominated by the contribution of a single energy eigenstate. 

\vskip10pt
\noindent 
Secondly, of course, a generic $T(z)$ does not correspond to a primary state, in fact in general it will be a generic element of the Virasoro orbit associated to a primary state. In principle all states within a Virasoro orbit are connected to each other by conformal transformations. So it remains to determine which primary state is associated to the conformal family that our heavy state $|V\rangle$ actually belongs to.  This primary state should be related to our heavy state by some judicially chosen single-valued conformal transformation \cite{deBoer:2016bov}. The Fuchs equation mirrors this picture; the conjugacy class of the monodromy matrix around the unit circle remains invariant under the group of orientation preserving diffeomorphisms of the circle $\textrm{Diff}_0(S^1)$ \cite{Balog:1997zz}. This is due to the eigenvalue being related to the scaling weight of the primary state and is hence an orbit invariant. We will now construct an element of $\textrm{Diff}_0(S^1)$ similar to the one in \cite{Balog:1997zz} that brings us to the stress-energy tensor of a single primary operator inserted at the origin with a specific scaling dimension. We can connect to the intuition above by interpreting the diffeomorphism as the transformation that undoes the boundary gravitons. The Fuchs equation is given by

$$
\psi''(z)+\frac{6}{c}T(z)\psi(z)=0.
$$
We assume that $T(z)$ is such that the monodromy along the unit circle of solutions falls in the hyperbolic class corresponding to real eigenvalues with $|\lambda|>1$. A complete classification of the monodromy matrices is presented in Table \ref{orbits}. Consider the eigenbasis of solutions under the monodromy transformation along the unit circle
$$
\psi_1\left(e^{2\pi i}z\right)=\lambda \psi_1(z), \hspace{1cm} \psi_2\left(e^{2\pi i}z\right)=\frac{1}{\lambda} \psi_2(z).
$$
In this case it is clear that the ratio of these two solutions $f(z)=\psi_1(z)/\psi_2(z)$ has the following transformation property
$$
f\left( e^{2\pi i}z\right)=\lambda^2 f(z).
$$
From this transformation rule we can construct a function that is inherently single-valued on the unit circle
$$
u(z)=e^{\frac{i\pi}{\g}\log(f(z))}=f(z)^{\frac{i\pi}{\g}},
$$
since this function simply adds a term $2\pi i$ to the exponent after making a full circle. Here $\g$ is defined by $\g=\log(|\lambda|)$. To prove that the function above is an element of $\textrm{Diff}_0(S^1)$ we need to show further that it is smooth everywhere on the unit circle.  The assumption that our stress-energy tensor is an element of a hyperbolic orbit and that the zero-mode of the stress-energy tensor is bounded from below restricts the orbit to the coadjoint orbit $B_0(b)$, as per the classification given in \cite{Balog:1997zz,deBoer:2016bov}. The complete classification is briefly summarised in Table \ref{orbits}. It is known that if $T(z)$ is contained within $B_{0}(b)$, the eigenbasis of solutions to the Fuchs equations has no roots along the unit circle \cite{Garbarz:2014kaa,Sheikh-Jabbari:2016unm}. This implies that the ratio of solutions $f(z)$ is forced to be smooth on the unit circle. Therefore we can conclude that $u(z)\in \textrm{Diff}_0(S^1)$. Next, with some algebra we will show that this function, $u(z)$ has a very special and desirable property. 

Taking the Schwarzian derivative of $u(z)$ and making use of the group structure of the Schwarzian derivative gives 
\begin{equation}
S[u(f(z)),z]=S[f,z]+\left(\frac{df}{dz}\right)^2S[u,f].
\label{chainrule}
\end{equation}
From the theory of Fuchs equations we know that under the Wronskian normalization condition $\psi_2(z)\psi_1'(z)-\psi_1(z)\psi_2'(z)=1$ the following identity holds
$$
S[f,z]=\frac{12}{c}T(z),
$$
in fact it is the inverted version of the Schwarzian ODE we started with. Inserting this identity in \eqref{chainrule} and rearranging the terms a little bit yields
\begin{eqnarray*} 
T(z)&=&\left(\frac{du}{dz}\right)^2 \frac{c}{24\pi^2}\left(\g^2+\pi^2\right) u(z)^{-2} + \frac{c}{12}S[u,z] \nonumber \\
&\equiv&\left(\frac{du}{dz}\right)^2 {\tilde H} \, u(z)^{-2} + \frac{c}{12}S[u,z]
\end{eqnarray*}
This is, therefore, exactly the coordinate transformation that yields a uniform stress-tensor component of the form ${\tilde H}/u^2$, where the role of the scaling dimension is now played by
\begin{equation}
{\tilde H}\leftrightarrow \frac{c}{24\pi^2}\left(\g^2+\pi^2\right),
\label{Hconnection}
\end{equation}
Note that the scaling dimension has a minimum value given by $c/24$, this is enforced by the initial assumption that we are in a hyperbolic monodromy class, the minimum value correponds to the zero mass BTZ state in the bulk. Given the primary stress-energy tensor the monodromy matrix can be computed by explicitly calculating the path-ordered integral \cite{Ammon:2013hba}. But the associated Fuchs equation can also be solved explicitly. In fact in \cite{Fitzpatrick:2015zha} the authors found uniformized correlator associated to these solutions and find that they produce a thermal correlator with temperature $T_H=\frac{1}{2\pi}\sqrt{24H/c-1}$, inserting \eqref{Hconnection} and reinstating $\lambda$ through $\g=\log(|\lambda|)$ gives us the relationship
\begin{equation} \nonumber
|\lambda|=e^{2\pi^2 T_{H}},
\end{equation}
quoted in the previous section.

\section{Chern-Simons interpretation of the monodromy} \label{sec4}
As shown in the previous section, the late-time behavior of the probe correlation function is controlled by a monodromy matrix. In this section we will demonstrate that this matrix has a natural manifestation in the Chern-Simons formulation of $3d$ gravity. The discussion in this section is heavily based on \cite{Ammon:2013hba}, additional references include \cite{deBoer:2016bov,Ammon:2012wc,Fitzpatrick:2016mtp,deBoer:2014sna}. The topological nature of $3d$ gravity is most clearly expressed in terms of so(2,2) connection field $A$ \cite{Witten:1989sx}, 
$$
{\bf A}_i=e_i^{a}{\bf P}_a+ \omega_i^{a}{\bf J}_a,
$$
here $e_i^{a}$ and $\omega_i^a$ are respectively the vielbein and the spin connection associated to the Einstein-Hilbert action.  $P_a$ and $J_a$ denote generators for translation and Lorentz transformations respectively. After imposing that $A$ transforms as a non-abelian gauge field under local $SO(2,2)$ transformations it can be shown that the Chern-Simons action
\begin{equation} \nonumber
S_{CS}\left[\bf A\right]=\frac{k}{4\pi} \int \textrm{Tr}\left({\bf A}\wedge d{\bf A}+\frac{2}{3}{\bf A}\wedge {\bf A}\wedge {\bf A}\right),
\end{equation}
is equivalent to the Einstein-Hilbert action with negative cosmological constant. The constant $k$ is the level of the Chern-Simons theory. By means of the Brown-Henneaux formula it can be related to the central charge through $k=\frac{c}{6}$. It turns out to be convenient sometimes to use the decomposition of the gauge group $SO(2,2)$ as $SL(2,\mathbb{R})\times SL(2,\mathbb{R})$. The connection field ${\bf A}$, accordingly, also assumes a decomposition in terms of a pair of $SL(2,\mathbb{R})$ connections, $\{A, {\bar A}\}$ defined as
$$ A_i =  \left(\omega_i^{a} + \frac{1}{R} e_i^{a}\right)J^{(+)}_a $$  $$ {\bar A}_i =  \left(\omega_i^{a} - \frac{1}{R} e_i^{a}\right)J^{(-)}_a$$
where $J^{(\pm)}_a = \frac{1}{2}\left(J_a \pm R P_a\right)$. With this decomposition the Einstein Hilbert action takes the form 
$$
S_{\text{EH}} = S_{CS}\left[A\right] - S_{CS}\left[{\bar A}\right],
$$
and the constant, $R$ gets a natural interpretation as the radius of the AdS$_3$.

\subsection{Wilson loops in Chern-Simons theory}
In \cite{Ammon:2013hba} Wilson lines were considered as geometrical probes for the bulk gravitational theory which act as the Chern-Simons analogue of massive probe particles traveling along geodesics in the bulk geometry. To make this statement quantitative
\begin{equation} \nonumber
W_{\mathcal{R}}(C)\sim e^{-m L},
\end{equation}
the $\sim$ designates that this is an on-shell relation. Here $C$ designates the curve along which the Wilson line is defined on the boundary of AdS$_3$. $m$ is the mass of the relevant probe and $L$ as the proper length of the geodesic connecting the endpoints of the Wilson line supported at the boundary. The subscript $\mathcal{R}$ denotes the $SL(2,\mathbb{R})$ representation in which the Wilson line falls.  In order to allow for a continuous mass parameter for the probe the representation needs to be infinite dimensional, e.g. the highest weight representation of  $SL(2,\mathbb{R})$ \cite{Ammon:2013hba}. As we will find, the relation to the geodesic length, $L$ stated above will turn out to be extremely useful. Given an appropriate bulk geodesic, this length will be interpreted as the horizon of the diffeomorphism-equivalent black hole which in turn is to its temperature by means of the Bekenstein-hawking formula.

\vskip10pt
\noindent
The key point of \cite{Ammon:2013hba} is that the expectation value of the Wilson line is given by the path integral over the following action
$$
S(U,P,A,\bar{A})_C=\int_C ds \, \textrm{Tr}(PU^{-1}D_sU)+\lambda(s)\left(\textrm{Tr}(P^2)-\frac{1}{2}m^2\right).
$$
The Hilbert space of the quantum system described by the auxiliary field U and its conjugate momentum P corresponds to the vector space given by the carrier space of an infinite dimensional representation of $SL(2,\mathbb{R})$.  $s$ parametrizes the curve, $C$ and $D_s$ denotes covariant derivative with respect to $s$, defined as
$$
D_sU = \frac{dU}{ds} + A_s U - U {\bar A}_s, \,\,\,\,\,\;\;\;\;\;\;\;\;\;  A_s \equiv A_{\mu} \frac{dx^\mu}{ds}.
$$
We can choose the field $U$ to live in the fundamental representation, this way we will connect to our monodromy matrix. Note that we are only integrating over the auxiliary fields and consider the Chern-Simons field $A$ as a background field which fixes the background geometry. As a consequence, the probe mass can take any real positive value without affecting the dynamics. The equations of motion of this action are
$$
U^{-1}D_sU+2\lambda(s)P=0, \hspace{1cm} \partial_s P+\left[\bar{A}_s,P\right]=0,
$$
while the Lagrange multiplier $\lambda(s)$ imposes the constraint
$$
\textrm{Tr}(P^2)=\frac{1}{2}m^2.
$$
On shell, the action reduces to a very simple form
$$
S_{\textrm{on-shell}}=-m^2\int_C ds\, \lambda(s).
$$
When solving the equations of motion it is natural to define the function $\alpha(s)$ through $d\alpha/ds=\lambda(s)$. In this case the action simply becomes
$$
S_{\textrm{on-shell}}=-m^2\Delta \alpha,
$$
the point is now clear, we need to solve for $\Delta \alpha$ and equate it to $L/m$. Note that $\Delta \alpha$ only depends on the endpoints of the curve. We will show below that knowledge of the boundary conditions is sufficient to solve for $\Delta \alpha$.

\subsection{The nothingness trick}

We will here follow steps very similar to those presented in \cite{Ammon:2013hba}. The equations of motion are simple to solve in the `empty gauge', i.e. $A=\bar{A}=0$, so we will solve the equations of motion in that particular gauge and then gauge transform the resulting solution to generate more solutions. One could ask, how many solutions are gauge connected to the empty gauge? The answer is that the bulk equation of motion of Chern-Simons theory imposes that $A$ is locally flat, hence all solutions to the Einstein field equations have an associated connection field $A$ that is gauge connected to the empty gauge. These gauge transformations will, however, be multivalued in general.
\vskip10pt
\noindent In the empty gauge the equations of motion are given by
$$
U_0(s)^{-1} \partial_s U_0(s)+2\lambda(s)P_0(s)=0, \hspace{1cm} \partial_s P_0=0.
$$
These equations are easily solved, their general solutions are given by
$$
U_0(s)=u_0e^{-2\alpha(s) P_0},\hspace{1cm} P_0(s)=P_0,
$$
here as before $d\alpha/ds=\lambda(s)$. Under $SL(2,\mathbb{R})\times SL(2,\mathbb{R})$ gauge transformations the auxiliary field variables transform as 
\begin{equation}
U(s)\rightarrow L(s)U(s)R(s), \hspace{1cm} P(s)\rightarrow R^{-1}(s)P(s)R(s),
\label{gaugetrafos1}
\end{equation}
where $L(s)$ and $R(s)$ are elements of the fundamental representation of $SL(2,\mathbb{R})$. In order to obtain the gauge transformation that brings us to the required connection, we need to find a transformation $L$ and $R$ such that 
$$
LdL^{-1}=A=b^{-1}ab+b^{-1}db=b^{-1}
\begin{pmatrix}
0&1\\
-\frac{6}{c}T(z)&0\\
\end{pmatrix}b+b^{-1}db,
$$
and similarly
$$
R^{-1}dR=\bar{A}=b^{-1}
\begin{pmatrix}
0&1\\
-\frac{6}{c}\bar{T}(\bar{z})&0\\
\end{pmatrix}b+b^{-1}db.
$$
Here $a$ is the boundary connection defined as $a=\left(J_{-1}+\frac{6}{c}T(z)J_{1}\right)dz$. It is related to the full connection through $A=b^{-1}(a+d)b$, with $b=e^{\rho J_0}$. $J_i$ denote $SL(2,\mathbb{R})$ generators with $i = \{1,0,-1\}$. With some simple algebra it can be shown that a solution to the differential equation for $L$ is given by
\be
L=e^{-\rho J_0}{\cal P}\left\{ e^{-\int_0^{s_f} a(z) ds}\right\}.
\label{Lsol}
\ee
One can similarly demonstrate that
\be
R={\cal P}\left\{e^{\int_0^{s_f} \bar{a}({\bar z})ds}\right\}e^{-\rho J_0}.
\label{Rsol}
\ee
We know what gauge transformation we want, the trick will be to apply it to the boundary conditions of the field variables. We are interested in Wilson loops (i.e. we want to find a geodesic that closes in on itself), therefore we have to impose periodic boundary conditions to the field variables
$$
U(0)=U(s_f), \hspace{1cm} P(0)=P(s_f).
$$
Using the gauge transformation \eqref{gaugetrafos1} the boundary condition for $P$ gives us 
\begin{equation}
\left[ P_0,R(s_f)R^{-1}(0)\right]=0.
\label{PBC}
\end{equation}
This condition implies that $P_0$ commutes with the path-ordered integral over the right-moving boundary gauge component. Similarly working out the boundary condition on $U$ gives us
\begin{equation}
u_0^{-1}L(s_f)^{-1}L(0)u_0R(0)R(s_f)^{-1},
\label{UBC}
\end{equation}
where as an intermediate step we used the fact above that $P_0$ commutes with the product $R(s_f)R^{-1}(0)$. Plugging \eqref{Lsol} and \eqref{Rsol} in \eqref{UBC} yields
\begin{equation}
e^{-2\Delta \alpha P_0}=u_0^{-1}Mu_0\bar{M}^{-1},
\label{stuff}
\end{equation}
Here $M$ is the same matrix as the monodromy matrix $M$ defined in \eqref{pathorderedintegral} in section 3. $\bar{M}$ is defined in an analogous way as the path-ordered integral over $\bar{A}$. Due to our restriction to scalar heavy operators (i.e. $\bar{h}=h$) the barred path-ordered integral gives
$$
\bar{M}=\mathcal{P}\left\{\textrm{exp}(\oint \bar{a}\left(\bar{T}(\bar{z})\right)ds)\right\}=\mathcal{P}\left\{\textrm{exp}(\oint a\left(T(\bar{z})\right)ds\right\}=M^{-1},
$$
the physical reason for this being that from the perspective of the anti-holomorphic variable $\bar{z}$ the contour is followed with the opposite orientation, hence
\begin{equation}\nonumber
e^{-2\Delta \alpha P_0}=u_0^{-1}Mu_0M.
\end{equation}
We can use \eqref{PBC} to simultaneously diagonalize the left- and right-hand side. Denoting the matrix that diagonalizes them by $V$,
$$
e^{-2\Delta \alpha P_0}=\left(u_0V\right)^{-1}M\left(u_0V\right)\bar{M}^{-1}.
$$
As pointed out in \cite{Ammon:2013hba}, consistency between the left- and the right-hand side demands that $\left(u_0V\right)^{-1}M\left(u_0V\right)$ is a diagonal matrix. Therefore we can make the above expression cleaner by defining $\lambda_M$ to be the diagonal matrix whose components are the eigenvalues of $M$
$$
e^{-2\Delta \alpha P_0}=\lambda_M^2.
$$
We can take the matrix logarithm on both sides, since all matrices involved are diagonal this reduces to the logarithms of the components
$$
-\Delta \alpha P_0=\log(\lambda_M)=
\begin{pmatrix}
\log(|\lambda|)&0\\
0&-\log(|\lambda|)\\
\end{pmatrix}.
$$
We know $\textrm{Tr}(P_0)=0$ because it has to live in the Lie algebra of $SL(2,\mathbb{R})$. Then from the constraint $\textrm{Tr}(P_0^2)=\frac{1}{2}m^2$, the eigenvalues of $P_0$  are given by $\pm m/2$. Contracting both sides with the matrix
$$
J_0=
\begin{pmatrix}
\frac{1}{2}&0\\
0&-\frac{1}{2}\\
\end{pmatrix}
$$
gives us
$$
-\frac{m}{2}\Delta \alpha =\log(|\lambda|).
$$
Inserting the expression $\Delta \alpha = -L/m$ yields
$$
\log(|\lambda|)=\frac{1}{2}L.
$$
As mentioned before, we assume that this $L$ is the proper length of the geodesic connecting the endpoints of the Wilson loop. Due to the topological nature of 3d gravity we can perform gauge transformations that correspond to continuously deforming the contour, the argument is that if there is a black hole in $AdS_3$ we can only continuously deform a contour to a minimum size measured by the holonomy. Hence $L$ measures the proper size of the horizon of black hole in the center of $AdS_3$. Making use of the Bekenstein-Hawking formule $S=A/4G$ and the Brown-Henneaux formula $c=\frac{3}{2G}$ finally gives us
$$
\log(|\lambda|)=2\pi^2 T_H.
$$
This is the relationship between the eigenvalues of the monodromy matrix $M$ and the temperature of the final state bulk black hole claimed in \eqref{scalinglaw}.

\section{An example and numerical black holes with soft gravitational hair}\label{numerics}
In the preceding section it was established that the eigenvalues of a certain monodromy matrix govern the late Lorentzian time behavior of probe correlation functions. In this section we will first consider an example, which is the smooth limit considered in \cite{Anous:2016kss}, the reason being that for this particular stress-energy tensor the monodromy matrix can be computed explicitly. In particular we will be able to see the transition from conical defect to black hole.

Secondly we perform a numerical analysis of the monodromy problem for various operator distributions and demonstrate that there can exist a deficiency between the initial energy injected in the bulk  and the mass of the final black hole state.  We interpret this deficiency as a fraction of the initial energy being frozen out in boundary gravitons, conserved charges associated with the asymptotic Brown-Henneaux Virasoro algebra of AdS$_3$.

\subsection{The continuous limit as an example}
One particular state in which the monodromy matrix can be computed explicitly is the isotropically collapsing shell of null dust constructed in \cite{Anous:2016kss}. This smooth limit for the initial and final states  was studied in \cite{Anous:2016kss}, where the number of insertions is taken to infinity, which leads to a simpler expression of the stress tensor. It is an example where the thermalization of a Lorentzian correlator can be shown explicitly. It closely resembles an eternal black hole state. The state on which the correlator is evaluated is given by
$$
|V\rangle= \lim_{n\rightarrow\infty}\frac{1}{\mathcal{N}}\prod_{k=1}^{n}O(e_k)|0\rangle, \hspace{5mm} e_k=(1-\sigma)e^{2\pi i(k-1)/n}.
$$
In the aforementioned paper it was shown that the stress-energy wave function for this particular state can be written as
$$
T(z)=\frac{K}{z^2}\Theta(|z|-1+\sigma)\theta(1-|z|+\sigma),
$$
where $K$ is a constant related to the CFT data through $K=\frac{H}{\sigma}$ and $\theta(z)$ denotes the heavyside function. It can easily be checked that this state has no gravitational hair, the reason being that along the unit circle stress-energy tensor takes the form of the lowest energy element of either a $\mathcal{B}_{0}(b)$ or $\mathcal{C}_{\nu}$ Virasoro coadjoint orbit\cite{Balog:1997zz}. The stress tensor above is very similar to the one of an eternal black hole, the difference is the compact support of the above stress-energy tensor.  For any cycle contained within the annulus the only visible pole would be the double pole located at $z=0$, for this reason it is possible to explicitly solve the path-ordered integral in \eqref{pathorderedintegral} in a straightforward manner. Inserting this stress-energy wave function into the flat connection \eqref{connection} gives
$$
a(z)=
\begin{pmatrix}
0&1\\
-\frac{6K}{c \,z^2}&0\\
\end{pmatrix}.
$$
The relevant information contained in the monodromy matrix are its eigenvalues. Hence we are only interested in its conjugacy class. We only need the connection up to an $\textrm{SL}(2,\mathbb{C})$ gauge transformation. There exists a theorem  that states that there always exists a gauge transformation such that around a singular point $z_0$ of $A(z)$ the connection can be transformed to a new connection $a(z)\rightarrow (z-z_0)^{-1-R_0}a_0(z)$, where the matrix elements of $a_0(z)$ are all regular functions of $z$ \cite{Castro:2013lba} \cite{roehrl1962}. The integer power $R_0$ is the Poincar\'e rank of the singular point, $R_0=0$ corresponds to a regular singular point. The connection transforms under a gauge transformation in the usual way
$$
a(z)\rightarrow U a U^{-1} + \left( \frac{d}{dz} U\right) U^{-1},
$$
a simple proposal for a gauge transformation that brings the connection to minimal form is the matrix that locally diagonalizes the connection A(z), specifically
$$
U(z)=
\begin{pmatrix}
-\frac{\sqrt{-\frac{3K}{2c}}}{z}&\frac{1}{2}\\
\frac{\sqrt{-\frac{3K}{2c}}}{z}&\frac{1}{2}\\
\end{pmatrix},
$$
this matrix is the inverse of the matrix of eigenvectors of $a(z)$. Under this transformation it can be found that the irregular part in $z$ (and in fact all $z$-dependence) factors out of the connection. The resulting path-ordered integral can be resolved by means of the following identity\cite{Castro:2013lba}
$$
M= P\left\{ e^{\oint dz \, \frac{1}{z}a_0(z)}\right\} =e^{2\pi i a_0(0)}.
$$
The heuristic argument is that one can continuously deform the contour to a limiting circle around a pole, at which point the diagonalizing matrix becomes exactly constant, subsequently rendering the path-ordered integral into a normal matrix integral. Unfortunately this particular approach does not generalize to multiple poles in a straight-forward manner, since one would need to consider integrating along paths connecting the poles. The eigenvalues of $2\pi i a_0(0)$ are given by
$$
\{ \lambda \} = i\pi \pm \pi \sqrt{\frac{24K}{c}-1},
$$
from which we can conclude that 
$$
\{ \lambda \}_{M} = -e^{\pm \pi\sqrt{\frac{24K}{c}-1}}.
$$
In \cite{Anous:2016kss} it was established that $T_{BH}=\frac{1}{2\pi}\sqrt{24K/c-1}$ therefore these eigenvalues are of the form derived in the previous sections. Which is exactly what was predicted above. One can observe both the the BTZ-mass threshold as well as the thermal behavior of the correlator at late time. This result is consistent with the statement made before that exact knowledge of the solutions of the Fuchs equation is over excessive for the purpose of finding the thermodynamic properties of the black hole at late time. Note that the analysis would have been identical for a state created by acting with a single primary operator at the origin on the vacuum.



\subsection{Numerical results including soft gravitational hair}
Solving the path-ordered integral in a general set-up is as mentioned in the previous section generically intractable. But since the path-orderd integral is the formal solution of an initial-value problem it is possible to draw some conclusions from numerical integration. 

One potentially confusing aspect we will focus on is that there can be very large deviations between the intial energy injected into the system and the mass of the final black hole state. We denote by $T_{H}'$ the temperature of a black hole in the hypothetical situation that all initial energy is converted into black hole rest mass. Using the Bekenstein Hawking formula for a non-rotational black hole \cite{Aharony:1999ti}
$$
S = \frac{2}{3}\pi^2c T_H',
$$ 
one can relate the temperature of the black hole to the Lorentzian energy $T_H  = \frac{\sqrt{3}}{\pi} \sqrt{E_L/c}$, where the Lorentzian energy on the cylinder is given by
$$
\frac{1}{2}E_L = \frac{1}{2\pi i}\left(-i\frac{\pi c}{12}+ \oint dz \, z T(z) \right),
$$ 
where the first term is the Casimir energy of the cylinder, which is also the minimal mass of the black hole \cite{Aharony:1999ti}. The factor 1/2 on the left-hand side is to emphasize that one has to add the identical holomorphic and anti-holomorphic contributions. Combining the above with the expression for the eigenvalues \ref{scalinglaw} one finds that these are given by 
$$
|\lambda'_\pm| = e^{\pm 2 \sqrt{3}\pi\sqrt{\frac{1}{i\pi c}\left(\oint dz \, z T(z)\right)-\frac{1}{12}} },
$$
in the case that all initial energy is converted into black hole mass at late time. The eigenvalues are pure phase, when the contour integral gives a value smaller than the minimal black hole mass, degenerate at the transition point ($\lambda =1$), and real and greater than one in the black hole phase. This picture is misleading though as generally not all energy is converted to black hole mass. If sufficient energy is present to form a minimal black hole, it is not necessarily true that thermalisation has to take over at late time \cite{deBoer:2016bov}.
The irregular part of the stress tensor expectation value is partially fixed by conformal invariance. Given a state in the semi-classical limit created by acting with $n$ primary operators on the vacuum the stress-tensor is given by
\begin{equation}
T(z)=\sum_{j=0}^{n-1} \left( \frac{h_j}{(z-z_j)^2} +\frac{h_j}{(z-1/\bar{z}_j)^2} + \frac{x_j}{z-z_j}+\frac{y_j}{z-1/\bar{z}_j}\right)+\textrm{regular terms}.
\end{equation}\label{eq:tensor}
where $h_j$ are the scaling dimensions of the operators insertions and $x_j$ and $y_j$ are accessory parameters that generally depend on the dynamical data of the theory. To simplify the computations we will assume that the scaling dimensions of all operators are equal and real, so $h_j = h$.  The accessory parameters satisfy the following constraint equations to ensure that there is no pole at infinity
\begin{equation} \nonumber
\begin{aligned}
 2h + x_j z_j + \frac{\bar{y}_j}{z_j} &= 0\\
\sum_{j=0}^{n-1}(x_j + y_j) &= 0 \\
\sum_{j=0}^{n-1} \text{Im}(x_j z_j + \frac{\bar{y}_j}{z_i}) &= 0.
\end{aligned}
\end{equation}
As mentioned in a previous section, in \cite{Hulik:2016ifr} it was shown that these constraints imply the reflection symmetry $z^2T(z)=\frac{1}{z^2}\bar{T}(1/z)$. 

In the following section we will assume that the insertions are evenly distributed around the circle. Without loss of generality we take $|z| = 1-\sigma$. In this case the $2n$ accessory parameters $x_j$ and $y_j$ can be expressed in terms of one free accessory parameter $x$ as follows
\begin{equation} \nonumber
\begin{aligned}
x_j &= e^{-\frac{j 2 \pi i}{n}}x \\
y_j &= -\left(2h +x(1-\sigma)\right)(1-\sigma)e^{-\frac{j 2 \pi i}{n}}.
\end{aligned}
\end{equation}

\subsubsection{Colliding $n$ clusters to form a black hole}
There is interesting physics that can be studied when we move away from the smooth limit, which is analytically accessible. Therefore in this section we present a short numerical study of a set of such states. We will assume identical pairs of operators that are evenly distributed along the unit circle.
\begin{figure*}[!htb]
	\centering
	\begin{subfigure}[b]{0.4\textwidth}
		\centering
		\includegraphics[width=.95\textwidth]{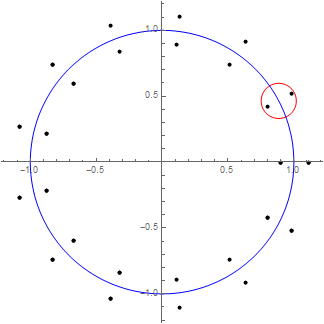}
		\caption{The set-up; Some insertions evenly distributed around the unit circle}
		\label{picturesetup}
	\end{subfigure}%
	~ 
	\begin{subfigure}[b]{0.6\textwidth}
		\centering
		\includegraphics[width=.95\textwidth]{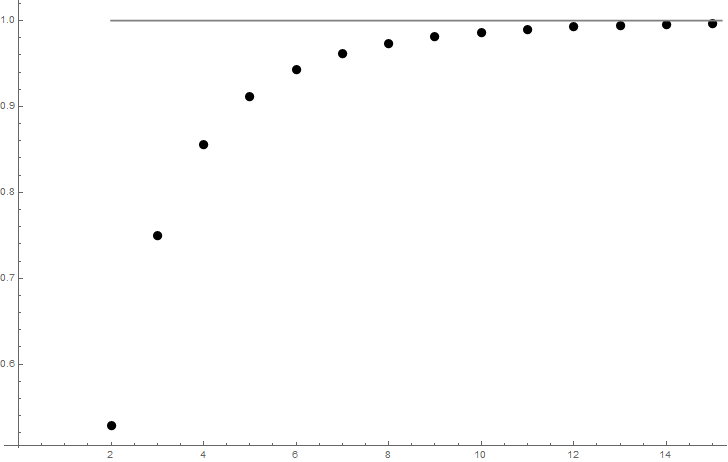}
		\caption{Fraction of no-hair black hole energy.}
		\label{pictures3}
	\end{subfigure}
	\caption{Colliding multiple conical defects.}
\end{figure*}

The free accessory parameter $x$ is fixed by demanding a trivial monodromy around a mirror pair of insertions $z_i$ and $1/\bar{z}_i$, depicted by the small red circle \ref{picturesetup}. Given the reflection property of the Fuchs equation, the following two monodromy problems are equivalent; either you fix your accessory parameters such that the monodromy around a mirror pair is trivial or you fix you accessory parameters such that any cycle around a single regular singular point yields a monodromy that falls within $SU(1,1)$. Given the reflection property of $T(z)$ it was shown by \cite{Hulik:2016ifr} that there exists a basis of solutions with the property that their ratio satisfies the same reflection property. From this, combined with the conjugacy class of $M$ being contained within $SU(1,1)$ it was shown that the monodromy of this pair of solutions around a mirror pair is trivial. Since the monodromy is a basis independent property this concluded the proof that the monodromy around a mirror pair is trivial. For an extensive proof of this statement we refer to \cite{Hulik:2016ifr}.
This demand fully determines the irregular part of the stress tensor and hence it is a well-defined problem to compute the monodromy around the unit circle. 
\noindent
Numerically integrating the connection field and determining its eigenvalue yields the mass of the black hole. In figure \ref{pictures3} one can see the fraction of energy stored in the black hole as a function of the number of regularly distributed insertions on a polygon, which was taken from two to fifteen. As the number of insertions increases the fraction goes to one, and it approaches the energy of the black hole that was created by a single primary acting at the origin. In this set-up the total energy $E\sim \frac{h n}{\sigma}$ that is injected was kept constant.
\vskip10pt
\noindent
For two colliding conical defects about half the energy is stored in boundary gravitons. Note that as the number of conical defects increases the amount of deficient energy decays rapidly. For the regularly distributed conical defects the resulting black hole is expected to be at rest in the Fefferman-Graham gauge. Note that we obtain the expected result that as we increase the number of conical defects we converge to the situation of rotational symmetry discussed in \cite{Anous:2016kss}.

\subsubsection{Non spherically symmetric distributions}
\begin{figure*}[!htb]
	\centering
	\begin{subfigure}[b]{0.5\textwidth}
		\centering
		\includegraphics[width=0.95\textwidth]{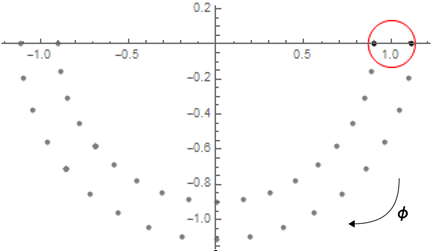}
		\subcaption{Collision angle}
		\label{setupangles}
	\end{subfigure}%
	 ~
	\begin{subfigure}[b]{0.5\textwidth}
		\centering
		\includegraphics[width=0.95\textwidth]{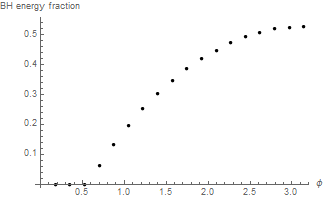}
		\subcaption{Resulting energy fraction}
		\label{resultangles}
	\end{subfigure}
	\caption{Colliding two conical defects under different angles.}
\end{figure*}

In figure \ref{setupangles} the set-up is displayed after relaxing the constraint of even distribution this will have the effect simulating oblique scattering. The first mirror pair is kept fixed at $\phi=0$ and a second pair is placed at an angle $\phi$, after which the eigenvalue is computed using the same procedure as described above. The eigenvalues were computed for angles varying from $\frac{\pi}{18}$ to $\pi$.
\vskip10pt
\noindent
In \ref{resultangles} the mass of the late time black hole, divided by the energy of the initial state is plotted for a collision angle varying from $\frac{\pi}{18}$ to $\pi$. It is expected that these asymmetrical states lead to an oscillating black hole in the bulk, partially explaining the small fraction of energy in the black hole. At the level of the algebra there is no distinction between the energy stored in momentum or the energy stored in other conserved spacetime symmetry charges that we call boundary gravitons. For small collision angles $\phi<\frac{\pi}{6}$ the black hole threshold is not exceeded. This could be interpret as two conical defects in stable orbits in the bulk, or a merger of two defects into one oscillating one.\\

\section{Discussion and outlook}
\noindent In this work we considered probe correlation functions on an arbitrary ``heavy state'' in 2d conformal field theory. In particular we studied the conditions under which the late time behavior of the vacuum block contribution of these correlators is indistinguishable from a thermal correlator. In the dual gravity picture this corresponds to the gravitational interaction of mass defects in AdS$_3$ which eventually, at a late Lorentzian time, collapses in a black hole spacetime, characterized by its Hawking temperature. In the  CFT$_2$ picture this heavy state was created through insertions of a small number of heavy primary operators whose conformal dimensions scale with the central charge, $c$ in the large $c$ limit. The distribution of the operator on the radial plane was kept arbitrary. 

\vskip10pt

\noindent For similar states created by a uniform distribution of a large ($n\rightarrow\infty$) number of operators, one can obtain systematic expansions of correlators of the form  $\mathcal{A} = \langle V|Q(z_1)Q(z_2) |V\rangle$ as a power series in $1\over c$ using monodromy methods for the dominant conformal block. The O(1) contribution corresponds to the semi-classical limit which at late Lorentzian time yields thermal correlators. The same conformal blocks can also be obtained for background states created by finite number of heavy operators using ``uniformization techniques". Although this technique works in principle for any finite number of operator insertions, it becomes a daunting task to explicitly find the uniformization coordinates as the number of operators increases. On the other hand in order to obtain a thermal collapse state dual to a black hole, one needs at least four heavy operators to create the initial heavy state. This technical obstacle is related to the fact that in order to obtain the uniformization coordinates, one needs to solve a differential equation of Fuchsian class. For four or more insertions these equations do not possess known explicit analytic solutions (except for some very specific choices of parameters which do not fit our particular need). 

\vskip10pt

\noindent {\it The main statement of our paper is that in order to extract the semi-classical thermal behavior of correlators at late Lorentzian time, one can actually bypass the requirement of solving the uniformization equation explicitly with non-trivial stress-energy function.} We note that the thermal behavior at late time is entirely controlled by the eigenvalue of the monodromy matrix along a curve on the unit circle. In this work we show that we can extract this eigenvalue even without the need to compute the full monodromy matrix for the solutions of Fuchs equation. {\it We present the explicit relation between this eigenvalue and the temperature of the final thermal state at late Lorentzian time. In the dual gravity picture this temperature corresponds to the Hawking temperature of the black hole formed through collisions among mass defects present in the bulk. This is the main result of this paper.}

\vskip10pt

\noindent We took two different routes to arrive at the aforementioned connection. The first proof relies on finding a conformal transformation that preserves the structure of the monodromy matrix while leading to a new Fuchs equation with a new stress energy function which can be solved analytically. For our Fuchsian system, we indeed found that one can construct an element of the orientation preserving diffeomorphism, $\textrm{Diff}_0(S^1)$ on the circle such that the new Fuchs equation we get is associated with a uniform stress energy function. In this form the Fuchs equation can be easily solved. However, since the diffeomorphism, $\textrm{Diff}_0(S^1)$ preserves the structure of the monodromy matrix, we can actually read off the required eigenvalue by studying this much simpler equation.
The second proof exploits the representation of the monodromy matrix in terms of a path ordered integral over a flat connection. Through the Chern-Simons formulation of pure gravity we relate this path ordered integral to the black hole surface area. Finally, upon using the Bekenstein-Hawking formula this finally yields the same expression for the eigenvalue.

\vskip10pt

\noindent However, having a Hawking temperature at the final state of the time evolution does not necessarily mean that one has a standard BTZ black hole at the end of the collapse process. In principle one can also have soft gravitational hair which can be interpreted as boundary gravitons. The bulk states associated with black holes dressed with gravitational hair correspond to points on the same Virasoro orbit. All these points are connected by large diffeomorphisms to a reference point identified with the bulk state describing the BTZ black hole. These are precisely the same diffeomorphisms we talked about in the first proof. The eigenvalue of the monodromy matrix is preserved on a particular Virasoro orbit. Following our identification of this eigenvalue with the temperature of the collapse state, this means all these bulk states related through diffeomorphisms must have the same temperature which in turn justifies the interpretation of these states as black holes states dressed with boundary gravitons. The way to distinguish between the BTZ black hole state and a state corresponding to black hole with gravitational hair is to compute the energy of the final configuration. The deficiency in the energy between the initial state and the predicted final black hole state without any gravitational hair is an indicator of the actual final state of the collapse process. In the work we set up a numerical program to demonstrate this phenomenon. We found that the energy stored in boundary gravitons can actually be sizable for states obtained by small number of heavy operator insertions. However, increasing the number of insertions seems to restore rotational symmetry leading to a stationary black hole with little or vanishing hair. This is consistent to our theoretical understanding that the reference point on the Virasoro orbit is associated with a uniform stress energy function in the large $c$ limit. We extended our numerical set up further to also include non-symmetric operator insertions  which additionally allows oscillating black holes as the final states of time evolution.

\vskip10pt

\noindent In this work we primarily focused on understanding the thermalization process starting from an atypical initial state. This thermal behavior is observed for the semiclassical correlators, namely from its leading order, O(1) contribution in the expansion in $1\over c$. It will be interesting to understand the $1\over c$ corrections to the correlator for which one would have to go beyond the vacuum block. These corrections are expected to modify the periodicity of Euclidean time order by order in $1\over c$ while still maintaining the KMS condition. This expectation follows from the dual gravity picture where these $1\over c$ corrections correspond to perturbation around the semiclassical large black hole solution. However, in \cite{Fitzpatrick:2015dlt} it was shown that this periodicity actually breaks down as a result of having another non-trivial monodromy which appears while going around the time circle. Although \cite{Fitzpatrick:2015dlt} argued that these new monodromies are unphysical and should go away with a resummation it will be interesting to understand this in our set up starting with an arbitrary initial state. It will also be worth investigating the role of boundary gravitons in the thermalization process in the next leading order of perturbative expansion. In particular, it will be interesting to understand the fate of the states corresponding to black holes with gravitational hair in higher order in $1\over c$. Another important and perhaps the most interesting direction of study will be to understand the eventual breakdown of the perturbative $1\over c$. This happens at the scrambling time \cite{Hayden:2007cs,Sekino:2008he,Asplund:2015eha} when the non-perturbative contributions $\sim e^{O(c)}$ become important. The associated timescale characterizes the onset of chaos \cite{Shenker:2013pqa,Maldacena:2015waa,Hosur:2015ylk,Caputa:2016tgt}. Work in this direction is in progress and we hope to report on this soon.

\section*{Acknowledgements}
We would like to thank K. Papadodimas specially for collaboration at an earlier stage of this work, numerous discussions at different stages of this project and for his valuable comments on the manuscript. We would also like to thank J. de Boer for useful discussions. The work of GV and JB are supported by the Royal Netherlands Academy of Arts and Sciences (KNAW). The work of SB is supported by the Knut and Alice Wallenberg Foundation under grant 113410212.

\appendix

\section{From Schwarzian equation to Fuchs equation}\label{appendix}

In this section it is shown how one can rewrite the Schwarzian type differential equation that was obtained by evaluating the $n$-point function in section 2 to a linear differential equation of Fuchsian class. We start with the highly non-linear differential equation
$$
\left(\frac{dz}{dw}\right)^{2}T(z(w))-\frac{c}{12}S[z(w),w]=0,
$$
where $S[z(w),w]$ is the Schwarzian derivative defined by
$$
S[z(w),w]\equiv\frac{z'''(w)}{z'(w)}-\frac{3}{2}\left(\frac{z''(w)}{z'(w)}\right)^2.
$$
As a first step we invert the differential equation, so that $T(z(w))$ is a function of the variable, instead of being a variable in the equation. This is done by means of the substitution $f(z(w))=\frac{dz(w)}{dw}=\frac{1}{w'(z)}$, where the prime denotes the derivative with respect to $w$. After some trivial manipulations one obtains
\begin{equation} \nonumber
\begin{aligned}
\frac{12}{c} T(z) +  \frac{f''}{f} - \frac{1}{2}\frac{( f')^2}{f^2}=0 .
\end{aligned}
\end{equation}
By adding and subtracting $\frac{1}{2}\frac{( f')^2}{f^2}$ we can simply replace the combination $\frac{ f''  }{f}-\frac{(f')^2}{f^2}$ by $\frac{d}{dz} \frac{f'(z)}{f(z)}$, this way we can obtain an equation that can be easily rewritten into a well known form
$$
\frac{12}{c} T(z) +  \frac{d}{dz}\frac{f'}{f} + \frac{1}{2}\frac{( f')^2}{f^2}=0.
$$
The previous expression can be simplified by defining the new function $g(z) = \frac{f(z)'}{f(z)}$. Substituting this into the expression above an equation of Riccati type is obtained,
$$
g'(z) + \half g(z)^2  + \frac{12}{c} T(z) =0.
$$
The Riccati type equation can be rewritten to a linear equation of Fuchsian type by defining $g(z) = 2 \frac{u'(z)}{u(z)}$ and substituting we obtain 
$$
u''(z)+\frac{6}{c}T(z)u(z) = 0.
$$
This demonstrates that the original Schwarzian equation used to define the uniformizing coordinate can be rewritten to a differential equation that is linear, which is a critical feature that will be exploited in section 3.

\section{Energy content of the model} \label{energyappendix}
In this section we provide some heuristic arguments to help obtain some scaling laws for the energy of the heavy states under consideration. These scaling laws provide us some intuition that we have applied in section 3 and 5 of the main body of the text. We will assume that if the separation of a mirror pair $\sigma$ is very small that the gravitational binding energy between mirror pairs is negligible in the gravitational dual in the weak gravity regime. Therefore we expect that the total energy of a heavy state is proportional to the sum of energies of the isolated mirror pairs. The stress tensor wavefunction of a single mirror pair is given by a conformal three-point function and can therefore be calculated explicitly 
$$
\langle O(1+\sigma)T(z)O(1-\sigma)\rangle=\frac{H}{\left((1-z)^2-\sigma^2\right)^2(2\sigma)^{2H-2}},
$$
note that we have implicitly assumed $\sigma <<1$. From this we find the expectation value
$$
\langle T(z)\rangle=\frac{\langle O(1+\sigma)T(z)O(1-\sigma)}{\langle O(1+\sigma)O(1-\sigma)}=\frac{4H\sigma^2}{\left((1-z)^2-\sigma^2\right)^2}.
$$
We see that the energy is peaked around $z=1$ with a value given by $\frac{4H}{\sigma^2}$, which drops of as $z^{-4}$ as $z$ goes to infinity, as we expect from a stress-energy tensor expectation value. The expression above will lead to the energy density of the state created by acting with the heavy operators. The following trick can be used to obtain the energy over the full Cauchy slice. First we map to the Euclidean cylinder
\begin{equation} \nonumber
\begin{aligned}
\langle O(1+\sigma)O(1-\sigma)\rangle &=e^{H\log(1+\sigma)+H\log(1-\sigma)}\langle O(\log(1+\sigma))O(\log(1-\sigma))\rangle\\ &\sim \langle O(\sigma)O(-\sigma)\rangle,
\end{aligned}
\end{equation}
where we dropped all terms of order $\mathcal{O}(\sigma^2)$. After Wick rotating to Lorentzian signature
\begin{equation} \nonumber
\langle O(i\sigma)O(-i\sigma)\rangle=\langle e^{-i (i\sigma) H}O e^{i (i\sigma) H} e^{-i (-i\sigma) H}O e^{i (-i\sigma) H}\rangle=\langle O e^{-2\sigma H}O\rangle,
\end{equation}
we can compute the total energy simply taking a derivative
$$
\frac{\langle OHO\rangle}{\langle OO\rangle}=\frac{-\frac{1}{2}\frac{d}{d\sigma}\langle O(i\sigma)O(-i\sigma)\rangle}{\langle O(i\sigma)O(-i\sigma)\rangle}=\frac{H}{\sigma}.
$$
This expression is similar to the one in \cite{Anous:2016kss}, note that it is only valid when $\sigma$ is small. We can simply take the second-derivative to find the expectation value of the Hamiltonian squared
$$
\frac{\langle OH^2 O\rangle}{\langle O O\rangle} =\frac{H(2H+1)}{2\sigma^2}.
$$
The standard deviation of the energy of the state is given by
$$
\Delta E=\sqrt{\langle E^2 \rangle - \langle E \rangle^2}=\sqrt{\frac{H}{2\sigma^2}}.
$$
This implies that our typical state has an energy variance that is wide enough for incorporating features like non-trivial quantum Poincar\'e  recurrences. While $\frac{\Delta E}{E}=\frac{1}{\sqrt{2 H}}$ is sufficiently small as to think of our resulting final state as a stationary black hole.

\section{Connection to the uniformization of the punctured Riemann sphere} \label{mathappendix}

There exists an interesting connection between the monodromy problem of the Fuchs equation and the classic uniformization problem of the psuedosphere that can be realised nicely in Liouville theory. In this appendix we will try to clarify this relation. The problem we have been studying is to construct CFT dual to a gravitational geometry in the bulk, containing some conical deficits. As mentioned before, computation of correlation functions in such a background  immensely simplifies by going to a coordinate system where the expectation value of the stress-energy tensor is zero. Holographically, this corresponds to creating a locally AdS spacetime through a non-trivial diffeomorphism. This, in a nutshell, is what we described as the ``uniformization problem''. 

This ``uniformization problem'' has a direct connection to Poincar\'e's classical uniformization theorem, which states that any punctured Riemann surface with genus, $g>1$ has a universal cover given by the unit disk (or equivalently the upper half-plane). In other words, this theorem proves the existence of a solution of our ``uniformization problem'', namely existence of the conformal transformation, $z \rightarrow w(z)$ mentioned before. Furthermore, the stress-energy tensor makes a natural appearance here when we cast the problem in the set up of Liouville field theory.

This interpretation of the theorem can be rephrased in a particular Riemann surface, namely the punctured disk with $N$ punctures ($\mathbb{C}\mathbb{P}^1/\{\zeta_i\}$) (as opposed to our Fuchs equation with 2$N$ regular singular points) corresponding to the insertion points, $\zeta_i$ of our CFT operators. The ``uniformization problem'' in this set up is to find the local Weyl factor $\phi({\zeta},\bar{\zeta})$ in order for the metric $ds^2=e^{\phi}d\zeta d\bar{\zeta}$ to have a constant negative curvature on the entire punctured disk. 
It turns out \cite{Takhtajan:2001uj} that the appropriate Weyl factor has the desired property on the punctured sphere if it satisfies the Liouville equation
$$
\partial_\zeta \partial_{\bar{\zeta}} \phi(\zeta,\bar{\zeta})-\frac{1}{2}e^{\phi(\zeta,\bar{\zeta})}=0,
$$
subject to the holomorphic boundary conditions 
$$
\phi(\zeta)=\left\{
\begin{matrix}
-2\log|\zeta-\zeta_i|-2\log|\log|\zeta-\zeta_i||+O(1) & \zeta\rightarrow \zeta_i\\
-2\log|\zeta|-2\log|\log|\zeta||+O(1) & |\zeta|\rightarrow 1
\end{matrix}\right.
$$
and similarly, for the anti-holomorphic part.
The first boundary conditions ensures that the metric is complete on the disk, the second condition makes sure that the punctured disk under consideration gets mapped to the Poincar\'e disk of area of $4\pi\left(N-1\right)$ \cite{Takhtajan:2001uj}. A holomorphic solution to the Liouville equation can be related to the holomorphic stress-energy tensor, $T(\zeta)$ by means of\footnote{note that there exists a sign convention difference between \cite{Hulik:2016ifr} and \cite{Takhtajan:2001uj}.} 
\begin{equation}
T(\zeta)=-\partial^2_{\zeta} \phi(\zeta)-\left(\partial_\zeta \phi(\zeta)\right)^2.
\label{LiouvilleEnergy}
\end{equation}
Similar expression also holds between their anti-holomorphic counterparts.

The only divergent point that $\phi(\zeta)$ is allowed to possess are near the punctures, since the boundary conditions restrict the divergences to be logarithmic it means that $T(\zeta)$ can only have at most second-order poles. Ignoring its unimportant regular part, the most general form of $T(\zeta)$ as a function of coordinates on the unit disk is given by
\begin{equation}
T(\zeta)=\sum_{i=1}^{N} \frac{\epsilon_i}{(\zeta-\zeta_i)^2}+\frac{c_i}{\zeta-\zeta_i},
\label{LiouvilleT}
\end{equation}
since the parameters $\epsilon_i$ control the strength of the singularities this whole expression is very similar to what appears as the stress-energy tensor wave function, $T(z)$ in our case. This precisely connects the uniformization problem we discussed in the text to the classical uniformization problem on the $N$-punctured Riemann surface. Therefore, the properties of the Liouville stress-energy tensor, $T(\zeta)$ can be attributed to the properties of our stress-energy wave function, $T(z)$.

In fact the general solution to the Liouville equation of motion is known, by simple substitution it can be shown that 
\begin{equation}
\phi(\zeta)=\log\left(\frac{4|f'(\zeta)|^2}{(1-|f|^2)^2}\right)=\log\left(\frac{4f'(\zeta)\bar{f}'(\bar{\zeta})}{(1-f(\zeta)\bar{f}(\bar{\zeta}))^2}\right)
\label{generalLiouville}
\end{equation}
solves the equation of motion, here $f(z)$ is generic meromorphic function. In fact by taking $f(z)=z$ you get the Weyl factor which corresponds to a metric of the unit disk with constant negative curvature (Poincar\'e disk)\cite{Zamolodchikov:2001ah}(i.e. in our context it would be a map from the unit disk to the unit disk). In order for the metric to be single-valued on the entire disk we need that the fundamental group of $f(z)$ on the disk to be a mapping of the form
\be \nonumber
M:\,\pi_1(\mathbb{C}\mathbb{P}^1/\{\zeta_i\})\rightarrow SU(1,1),
\ee
the restriction to $SU(1,1)$ as opposed to $SL(2,\mathbb{C})$ ensures the single-valuedness of the metric on the puntured disk. Furthermore by substituting \eqref{generalLiouville} into \eqref{LiouvilleEnergy} one can demonstrate that $f(z)$ satisfies the property
\be \nonumber
S[f(\zeta)]=2T(\zeta),
\ee
where $S[f(\zeta)]$ is the Schwartzian derivative of $f(\zeta)$ with respect to $z$. The real boundary condition of the function $f(\zeta)$ on the boundary of the unit disk invokes a reflection property of the resulting stress-energy tensor, a fact we exploit in section \ref{numerics}.

\bibliographystyle{JHEP}
\bibliography{bibliography}

\providecommand{\href}[2]{#2}\begingroup\raggedright\begin{thebibliography}{10}

\bibitem{Maldacena:1997re}
J.~M. Maldacena, {\it {The Large N limit of superconformal field theories and
  supergravity}},  {\em Adv.Theor.Math.Phys.} {\bf 2} (1998) 231--252,
  [\href{http://xxx.lanl.gov/abs/hep-th/9711200}{{\tt hep-th/9711200}}].

\bibitem{Banados:1992wn}
M.~Banados, C.~Teitelboim, and J.~Zanelli, {\it {The Black hole in
  three-dimensional space-time}},  {\em Phys. Rev. Lett.} {\bf 69} (1992)
  1849--1851, [\href{http://xxx.lanl.gov/abs/hep-th/9204099}{{\tt
  hep-th/9204099}}].

\bibitem{Maldacena:1998uz}
J.~M. Maldacena, J.~Michelson, and A.~Strominger, {\it {Anti-de Sitter
  fragmentation}},  {\em JHEP} {\bf 02} (1999) 011,
  [\href{http://xxx.lanl.gov/abs/hep-th/9812073}{{\tt hep-th/9812073}}].

\bibitem{Maldacena:2016hyu}
J.~Maldacena and D.~Stanford, {\it {Remarks on the Sachdev-Ye-Kitaev model}},
  {\em Phys. Rev.} {\bf D94} (2016), no.~10 106002,
  [\href{http://xxx.lanl.gov/abs/1604.0781}{{\tt arXiv:1604.0781}}].

\bibitem{Maldacena:2016upp}
J.~Maldacena, D.~Stanford, and Z.~Yang, {\it {Conformal symmetry and its
  breaking in two dimensional Nearly Anti-de-Sitter space}},  {\em PTEP} {\bf
  2016} (2016), no.~12 12C104, [\href{http://xxx.lanl.gov/abs/1606.0185}{{\tt
  arXiv:1606.0185}}].

\bibitem{srednicki1999approach}
M.~Srednicki, {\it The approach to thermal equilibrium in quantized chaotic
  systems},  {\em Journal of Physics A: Mathematical and General} {\bf 32}
  (1999), no.~7 1163.

\bibitem{Anous:2016kss}
T.~Anous, T.~Hartman, A.~Rovai, and J.~Sonner, {\it {Black Hole Collapse in the
  1/c Expansion}},  {\em JHEP} {\bf 07} (2016) 123,
  [\href{http://xxx.lanl.gov/abs/1603.0485}{{\tt arXiv:1603.0485}}].

\bibitem{Chen:2016dfb}
B.~Chen, J.-q. Wu, and J.-j. Zhang, {\it {Holographic Description of 2D
  Conformal Block in Semi-classical Limit}},  {\em JHEP} {\bf 10} (2016) 110,
  [\href{http://xxx.lanl.gov/abs/1609.0080}{{\tt arXiv:1609.0080}}].

\bibitem{Chen:2016kyz}
B.~Chen and J.-q. Wu, {\it {Holographic Entanglement Entropy For a Large Class
  of States in 2D CFT}},  {\em JHEP} {\bf 09} (2016) 015,
  [\href{http://xxx.lanl.gov/abs/1605.0675}{{\tt arXiv:1605.0675}}].

\bibitem{Fitzpatrick:2015zha}
A.~L. Fitzpatrick, J.~Kaplan, and M.~T. Walters, {\it {Virasoro Conformal
  Blocks and Thermality from Classical Background Fields}},  {\em JHEP} {\bf
  11} (2015) 200, [\href{http://xxx.lanl.gov/abs/1501.0531}{{\tt
  arXiv:1501.0531}}].

\bibitem{Fitzpatrick:2015dlt}
A.~L. Fitzpatrick and J.~Kaplan, {\it {Conformal Blocks Beyond the
  Semi-Classical Limit}},  {\em JHEP} {\bf 05} (2016) 075,
  [\href{http://xxx.lanl.gov/abs/1512.0305}{{\tt arXiv:1512.0305}}].

\bibitem{Hadasz:2006rb}
L.~Hadasz and Z.~Jaskolski, {\it {Liouville theory and uniformization of
  four-punctured sphere}},  {\em J. Math. Phys.} {\bf 47} (2006) 082304,
  [\href{http://xxx.lanl.gov/abs/hep-th/0604187}{{\tt hep-th/0604187}}].

\bibitem{Castro:2013lba}
A.~Castro, J.~M. Lapan, A.~Maloney, and M.~J. Rodriguez, {\it {Black Hole
  Scattering from Monodromy}},  {\em Class. Quant. Grav.} {\bf 30} (2013)
  165005, [\href{http://xxx.lanl.gov/abs/1304.3781}{{\tt arXiv:1304.3781}}].

\bibitem{Pasterski:2015tva}
S.~Pasterski, A.~Strominger, and A.~Zhiboedov, {\it {New Gravitational
  Memories}},  {\em JHEP} {\bf 12} (2016) 053,
  [\href{http://xxx.lanl.gov/abs/1502.0612}{{\tt arXiv:1502.0612}}].

\bibitem{Hawking:2016msc}
S.~W. Hawking, M.~J. Perry, and A.~Strominger, {\it {Soft Hair on Black
  Holes}},  {\em Phys. Rev. Lett.} {\bf 116} (2016), no.~23 231301,
  [\href{http://xxx.lanl.gov/abs/1601.0092}{{\tt arXiv:1601.0092}}].

\bibitem{Hawking:2016sgy}
S.~W. Hawking, M.~J. Perry, and A.~Strominger, {\it {Superrotation Charge and
  Supertranslation Hair on Black Holes}},  {\em JHEP} {\bf 05} (2017) 161,
  [\href{http://xxx.lanl.gov/abs/1611.0917}{{\tt arXiv:1611.0917}}].

\bibitem{Strominger:2014pwa}
A.~Strominger and A.~Zhiboedov, {\it {Gravitational Memory, BMS
  Supertranslations and Soft Theorems}},  {\em JHEP} {\bf 01} (2016) 086,
  [\href{http://xxx.lanl.gov/abs/1411.5745}{{\tt arXiv:1411.5745}}].

\bibitem{Hartman:2013mia}
T.~Hartman, {\it {Entanglement Entropy at Large Central Charge}},
  \href{http://xxx.lanl.gov/abs/1303.6955}{{\tt arXiv:1303.6955}}.

\bibitem{Hartman:2014oaa}
T.~Hartman, C.~A. Keller, and B.~Stoica, {\it {Universal Spectrum of 2d
  Conformal Field Theory in the Large c Limit}},  {\em JHEP} {\bf 09} (2014)
  118, [\href{http://xxx.lanl.gov/abs/1405.5137}{{\tt arXiv:1405.5137}}].

\bibitem{Heemskerk:2009pn}
I.~Heemskerk, J.~Penedones, J.~Polchinski, and J.~Sully, {\it {Holography from
  Conformal Field Theory}},  {\em JHEP} {\bf 10} (2009) 079,
  [\href{http://xxx.lanl.gov/abs/0907.0151}{{\tt arXiv:0907.0151}}].

\bibitem{Fitzpatrick:2010zm}
A.~L. Fitzpatrick, E.~Katz, D.~Poland, and D.~Simmons-Duffin, {\it {Effective
  Conformal Theory and the Flat-Space Limit of AdS}},  {\em JHEP} {\bf 07}
  (2011) 023, [\href{http://xxx.lanl.gov/abs/1007.2412}{{\tt
  arXiv:1007.2412}}].

\bibitem{ElShowk:2011ag}
S.~El-Showk and K.~Papadodimas, {\it {Emergent Spacetime and Holographic
  CFTs}},  \href{http://xxx.lanl.gov/abs/1101.4163}{{\tt arXiv:1101.4163}}.

\bibitem{Osterwalder:1974tc}
K.~Osterwalder and R.~Schrader, {\it {Axioms for Euclidean Green's Functions.
  2.}},  {\em Commun. Math. Phys.} {\bf 42} (1975) 281.

\bibitem{Luscher:1974ez}
M.~Luscher and G.~Mack, {\it {Global Conformal Invariance in Quantum Field
  Theory}},  {\em Commun. Math. Phys.} {\bf 41} (1975) 203--234.

\bibitem{Streater:1989vi}
R.~F. Streater and A.~S. Wightman, {\em {PCT, spin and statistics, and all
  that}}.
\newblock 1989.

\bibitem{Roberts:2014ifa}
D.~A. Roberts and D.~Stanford, {\it {Two-dimensional conformal field theory and
  the butterfly effect}},  {\em Phys. Rev. Lett.} {\bf 115} (2015), no.~13
  131603, [\href{http://xxx.lanl.gov/abs/1412.5123}{{\tt arXiv:1412.5123}}].

\bibitem{Hulik:2016ifr}
O.~Hulík, T.~Procházka, and J.~Raeymaekers, {\it {Multi-centered AdS$_{3}$
  solutions from Virasoro conformal blocks}},  {\em JHEP} {\bf 03} (2017) 129,
  [\href{http://xxx.lanl.gov/abs/1612.0387}{{\tt arXiv:1612.0387}}].

\bibitem{Martinec:1998wm}
E.~J. Martinec, {\it {Conformal field theory, geometry, and entropy}},
  \href{http://xxx.lanl.gov/abs/hep-th/9809021}{{\tt hep-th/9809021}}.

\bibitem{Witten:1987ty}
E.~Witten, {\it {Coadjoint Orbits of the Virasoro Group}},  {\em Commun. Math.
  Phys.} {\bf 114} (1988) 1.

\bibitem{Banados:1998gg}
M.~Banados, {\it {Three-dimensional quantum geometry and black holes}},  {\em
  AIP Conf. Proc.} {\bf 484} (1999), no.~1 147--169,
  [\href{http://xxx.lanl.gov/abs/hep-th/9901148}{{\tt hep-th/9901148}}].

\bibitem{Brown:1986nw}
J.~D. Brown and M.~Henneaux, {\it {Central Charges in the Canonical Realization
  of Asymptotic Symmetries: An Example from Three-Dimensional Gravity}},  {\em
  Commun. Math. Phys.} {\bf 104} (1986) 207--226.

\bibitem{Balog:1997zz}
J.~Balog, L.~Feher, and L.~Palla, {\it {Coadjoint orbits of the Virasoro
  algebra and the global Liouville equation}},  {\em Int. J. Mod. Phys.} {\bf
  A13} (1998) 315--362, [\href{http://xxx.lanl.gov/abs/hep-th/9703045}{{\tt
  hep-th/9703045}}].

\bibitem{Compere:2015knw}
G.~Compère, P.~Mao, A.~Seraj, and M.~M. Sheikh-Jabbari, {\it {Symplectic and
  Killing symmetries of AdS$_{3}$ gravity: holographic vs boundary gravitons}},
   {\em JHEP} {\bf 01} (2016) 080,
  [\href{http://xxx.lanl.gov/abs/1511.0607}{{\tt arXiv:1511.0607}}].

\bibitem{Garbarz:2014kaa}
A.~Garbarz and M.~Leston, {\it {Classification of Boundary Gravitons in AdS$_3$
  Gravity}},  {\em JHEP} {\bf 05} (2014) 141,
  [\href{http://xxx.lanl.gov/abs/1403.3367}{{\tt arXiv:1403.3367}}].

\bibitem{Sheikh-Jabbari:2016unm}
M.~M. Sheikh-Jabbari and H.~Yavartanoo, {\it {On 3d bulk geometry of Virasoro
  coadjoint orbits: orbit invariant charges and Virasoro hair on locally
  AdS$_3$ geometries}},  {\em Eur. Phys. J.} {\bf C76} (2016), no.~9 493,
  [\href{http://xxx.lanl.gov/abs/1603.0527}{{\tt arXiv:1603.0527}}].

\bibitem{deBoer:2016bov}
J.~de~Boer and D.~Engelhardt, {\it {Remarks on thermalization in 2D CFT}},
  {\em Phys. Rev.} {\bf D94} (2016), no.~12 126019,
  [\href{http://xxx.lanl.gov/abs/1604.0532}{{\tt arXiv:1604.0532}}].

\bibitem{Ammon:2013hba}
M.~Ammon, A.~Castro, and N.~Iqbal, {\it {Wilson Lines and Entanglement Entropy
  in Higher Spin Gravity}},  {\em JHEP} {\bf 10} (2013) 110,
  [\href{http://xxx.lanl.gov/abs/1306.4338}{{\tt arXiv:1306.4338}}].

\bibitem{Ammon:2012wc}
M.~Ammon, M.~Gutperle, P.~Kraus, and E.~Perlmutter, {\it {Black holes in three
  dimensional higher spin gravity: A review}},  {\em J. Phys.} {\bf A46} (2013)
  214001, [\href{http://xxx.lanl.gov/abs/1208.5182}{{\tt arXiv:1208.5182}}].

\bibitem{Fitzpatrick:2016mtp}
A.~L. Fitzpatrick, J.~Kaplan, D.~Li, and J.~Wang, {\it {Exact Virasoro Blocks
  from Wilson Lines and Background-Independent Operators}},  {\em JHEP} {\bf
  07} (2017) 092, [\href{http://xxx.lanl.gov/abs/1612.0638}{{\tt
  arXiv:1612.0638}}].

\bibitem{deBoer:2014sna}
J.~de~Boer, A.~Castro, E.~Hijano, J.~I. Jottar, and P.~Kraus, {\it {Higher spin
  entanglement and $ {\mathcal{W}}_{\mathrm{N}} $ conformal blocks}},  {\em
  JHEP} {\bf 07} (2015) 168, [\href{http://xxx.lanl.gov/abs/1412.7520}{{\tt
  arXiv:1412.7520}}].

\bibitem{Witten:1989sx}
E.~{Witten}, {\it {Topology-changing amplitudes in 2 + 1 dimensional gravity}},
   {\em Nuclear Physics B} {\bf 323} (Aug., 1989) 113--140.

\bibitem{roehrl1962}
H.~Röhrl, {\it Holomorphic fiber bundles over riemann surfaces},  {\em Bull.
  Amer. Math. Soc.} {\bf 68} (05, 1962) 125--160.

\bibitem{Aharony:1999ti}
O.~Aharony, S.~S. Gubser, J.~M. Maldacena, H.~Ooguri, and Y.~Oz, {\it {Large N
  field theories, string theory and gravity}},  {\em Phys. Rept.} {\bf 323}
  (2000) 183--386, [\href{http://xxx.lanl.gov/abs/hep-th/9905111}{{\tt
  hep-th/9905111}}].

\bibitem{Hayden:2007cs}
P.~Hayden and J.~Preskill, {\it {Black holes as mirrors: Quantum information in
  random subsystems}},  {\em JHEP} {\bf 09} (2007) 120,
  [\href{http://xxx.lanl.gov/abs/0708.4025}{{\tt arXiv:0708.4025}}].

\bibitem{Sekino:2008he}
Y.~Sekino and L.~Susskind, {\it {Fast Scramblers}},  {\em JHEP} {\bf 10} (2008)
  065, [\href{http://xxx.lanl.gov/abs/0808.2096}{{\tt arXiv:0808.2096}}].

\bibitem{Asplund:2015eha}
C.~T. Asplund, A.~Bernamonti, F.~Galli, and T.~Hartman, {\it {Entanglement
  Scrambling in 2d Conformal Field Theory}},  {\em JHEP} {\bf 09} (2015) 110,
  [\href{http://xxx.lanl.gov/abs/1506.0377}{{\tt arXiv:1506.0377}}].

\bibitem{Shenker:2013pqa}
S.~H. Shenker and D.~Stanford, {\it {Black holes and the butterfly effect}},
  {\em JHEP} {\bf 03} (2014) 067,
  [\href{http://xxx.lanl.gov/abs/1306.0622}{{\tt arXiv:1306.0622}}].

\bibitem{Maldacena:2015waa}
J.~Maldacena, S.~H. Shenker, and D.~Stanford, {\it {A bound on chaos}},  {\em
  JHEP} {\bf 08} (2016) 106, [\href{http://xxx.lanl.gov/abs/1503.0140}{{\tt
  arXiv:1503.0140}}].

\bibitem{Hosur:2015ylk}
P.~Hosur, X.-L. Qi, D.~A. Roberts, and B.~Yoshida, {\it {Chaos in quantum
  channels}},  {\em JHEP} {\bf 02} (2016) 004,
  [\href{http://xxx.lanl.gov/abs/1511.0402}{{\tt arXiv:1511.0402}}].

\bibitem{Caputa:2016tgt}
P.~Caputa, T.~Numasawa, and A.~Veliz-Osorio, {\it {Out-of-time-ordered
  correlators and purity in rational conformal field theories}},  {\em PTEP}
  {\bf 2016} (2016), no.~11 113B06,
  [\href{http://xxx.lanl.gov/abs/1602.0654}{{\tt arXiv:1602.0654}}].

\bibitem{Takhtajan:2001uj}
L.~Takhtajan and P.~Zograf, {\it {Hyperbolic 2 spheres with conical
  singularities, accessory parameters and Kahler metrics on $M(0,n)$}},
  \href{http://xxx.lanl.gov/abs/math/0112170}{{\tt math/0112170}}.

\bibitem{Zamolodchikov:2001ah}
A.~B. Zamolodchikov and A.~B. Zamolodchikov, {\it {Liouville field theory on a
  pseudosphere}},  \href{http://xxx.lanl.gov/abs/hep-th/0101152}{{\tt
  hep-th/0101152}}.

\end{thebibliography}\endgroup

\end{document}